\documentclass[vecphys,letterpaper]{svmult}

\multiply\textheight by 52\relax

\renewcommand{\index}[1]{{}}

\usepackage{graphicx}
\usepackage{multicol}
\usepackage[nosort]{cite}
\usepackage[bottom]{footmisc}
\usepackage{amsmath}
\usepackage{amsfonts}
\usepackage{amssymb}
\usepackage{bm}

\begin{document}

  \title{Dynamical Mean-Field Theory}\label{Avella:Vollhardt}
  \titlerunning{Dynamical Mean-Field Theory}
   \author{Dieter Vollhardt\inst{1} \and Krzysztof Byczuk\inst{2} \and Marcus Kollar\inst{1}}
  \authorrunning{Vollhardt, Byczuk, Kollar}
    \institute{Theoretical Physics III, Center for Electronic
    Correlations and Magnetism, Institute of Physics, University of
    Augsburg, 86135 Augsburg, Germany
    \and Institute of Theoretical Physics, Faculty of Physics,
     University of Warsaw, ul.\ Ho\.za~69, 00-681 Warszawa, Poland\\
     ~\\
     ~\\
     ~
   }
  \maketitle

  \begin{abstract}
    The dynamical mean-field theory\protect\index{dynamical mean-field theory} (DMFT)\protect\index{DMFT|see{dynamical mean-field theory}} is a widely applicable
    approximation scheme for the investigation of correlated quantum
    many-particle systems on a lattice, e.g., electrons in solids and
    cold atoms in optical lattices. In particular, the combination of
    the DMFT with conventional methods for the calculation of
    electronic band structures has led to a powerful numerical
    approach which allows one to explore the properties of correlated
    materials. In this introductory article we discuss the foundations
    of the DMFT, derive the underlying self-consistency equations, and
    present several applications which have provided important
    insights into the properties of correlated matter.
  \end{abstract}


  \section{Motivation}

\subsection{Electronic Correlations}

Already in 1937, at the outset of modern solid state physics, de  Boer and
Verwey \cite{Vollhardt:deboer} drew attention to the surprising  properties of
materials with incompletely filled 3$d$-bands.  This observation prompted
Mott and Peierls  \cite{Vollhardt:mott37} to  discuss the interaction between the
electrons. Ever since  transition metal oxides\index{transition metal oxides} (TMOs) were investigated
intensively \cite{Vollhardt:tokura}. It is now well-known that in many materials with
partially filled electron  shells, such as the 3$d$ transition metals V and
Ni and their  oxides, or 4$f$ rare-earth metals such as Ce, electrons
occupy narrow  orbitals. The spatial confinement enhances the effect of the
Coulomb interaction between the electrons, making them ``strongly
correlated''. Correlation effects
can lead to profound quantitative and  qualitative changes of the physical
properties of electronic systems  as compared to non-interacting particles.
In particular, they often  respond very strongly to changes in external
parameters. This is  expressed by large renormalizations of the response
functions of the  system, e.g., of the spin susceptibility and the charge
compressibility.  In particular, the interplay  between the spin, charge and
orbital degrees of freedom of the  correlated $d$ and $f$ electrons and with
the lattice degrees of  freedom leads to an  amazing multitude of ordering
phenomena  and other fascinating  properties, including high temperature
superconductivity, colossal  magnetoresistance and Mott metal-insulator
transitions  \cite{Vollhardt:tokura}.

\subsection{The Hubbard Model}

The simplest microscopic model describing interacting electrons in a solid
is  the one-band, spin-1/2 Hubbard model\index{Hubbard model} \cite{Vollhardt:Gutzwiller,Vollhardt:HubbardI,Vollhardt:Kanamori}
where the interaction between  the electrons is assumed to be so strongly
screened that it is purely local. More generally the Hubbard model applies
to lattice fermions with a point interaction, such as ultra-cold fermionic atoms\index{ultra-cold atomic gases}
in optical lattices\index{optical lattices} where the interaction is indeed extremely short ranged.
The Hamiltonian consists of two terms, the kinetic  energy $\hat{H}_0$ and
the interaction energy $\hat{H}_I$ (here and  in the following operators are
denoted by a hat):
\begin{subequations}
\label{Vollhardt:G11.7}
\begin{eqnarray}  \label{Vollhardt:G11.7a}
\hat{H} & = & \hat{H}_0 + \hat{H}_I \\[10pt]
\hat{H}_0 & = & \sum_{i , j} \sum_{\sigma} t_{ij} \hat{c}_{i
\sigma}^{+} \hat{c}_{j \sigma}^{} = \sum_{\bm{k} , \sigma} \epsilon_{\bm{k}}
\hat{n}_{\bm{k} \sigma}^{}  \label{Vollhardt:G11.7b} \\[10pt]
\hat{H}_{\mathrm{I}} & = & U \sum_{i} \hat{n}_{i \uparrow} \hat{n}_{i
\downarrow},  \label{Vollhardt:G11.7c}
\end{eqnarray}
where $\hat{c}_{i \sigma}^{+} (\hat{c}_{i \sigma}^{})$ are creation
(annihilation) operators of fermions with spin $\sigma$ at site  $\bm{R}_i$
(for simplicity denoted by $i$), and $\hat{n}_{i \sigma}^{} = \hat{c}_{i
\sigma}^{+}  \hat{c}_{i \sigma}^{}$. The Fourier transform of the kinetic
energy  in \eqref{Vollhardt:G11.7b}, where $t_{ij}$ is the amplitude for hopping
between sites $i$ and $j$, involves  the dispersion $\epsilon_{\bm{k}}$\index{dispersion relation} and
the momentum distribution  operator $\hat{n}_{\bm{k} \sigma}^{}$. This model
provides the basis for most of the theoretical research on  correlated
electrons during the last decades.

The Hubbard model describes an  interacting many-body system which cannot be
solved  analytically, except in dimension $d=1$ for nearest-neighbor hopping
\cite{Vollhardt:Lieb+Wu}. In the absence of exact solutions there is clearly a great
need for reliable, controlled  approximation schemes for this model.
However, such approximations are not easy to construct as the following
observation shows.
When viewed as a  function of time a given site of the lattice will
sometimes be empty, singly occupied  or doubly occupied. For strong
repulsion $U$ double  occupations are energetically very unfavorable and are
therefore  strongly suppressed, implying  $\langle \hat{n}_{i \uparrow}
\hat{n}_{i \downarrow} \rangle \neq  \langle \hat{n}_{i\uparrow} \rangle \langle
\hat{n}_{i\downarrow}  \rangle $. Therefore approximation schemes based on
the factorization of the interaction term,
e.g., Hartree-Fock-type\index{Hartree-Fock method} mean-field theories, are generally insufficient to
explain the physics of electrons in their  paramagnetic phase beyond the
limit of weak interactions. This is due to the fact that in such
approximations the  interaction is described only as an average, static potential,
whereby correlations, i.e., dynamical many-body effects due to the interaction of individual electrons,
are excluded from the
beginning. Hence correlation phenomena such as the Mott-Hubbard
metal-insulator transition cannot be described by such approximations. This
clearly shows the need for comprehensive approximation schemes, which are
applicable for all values of the input parameters, e.g., coupling parameters
and temperature, diagrammatically controlled, and thermodynamically
consistent \cite{Vollhardt:Vollhardt-Salerno}.

\subsection{Construction of Comprehensive Mean-Field Theories for
Many-Particle Models}

There exists a well-established branch of approximation techniques  which
makes use of the simplifications that occur when some  parameter is taken to
be large (in fact, infinite), e.g., the length  of the spins $S$, the spin
degeneracy $N$, the spatial dimension $d$, or the coordination number  $Z$,
i.e., the number of nearest neighbors of a lattice site.\footnote{The coordination number $Z$ is determined by the dimension $d$ and the
lattice structure. Already in $d = 3$ the coordination number can be quite large, e.g., $Z = 6$
for a simple cubic lattice, $Z = 8$ for a bcc lattice and $Z  = 12$ for an
fcc-lattice, making its inverse, $1/Z$, rather small. It is then natural to
consider the limit $Z \to  \infty$ to simplify the problem. For a hypercubic
lattice, obtained by generalizing the simple cubic lattice in $d=3$ to
arbitrary dimensions, one has $Z=2d$. The limit $d\rightarrow \infty $ is
then equivalent to $Z\rightarrow \infty $.
Several standard approximation schemes which are  commonly used to explain
experimental results in dimension $d = 3$  are exact only in $d,Z = \infty$
\cite{Vollhardt:Jerusalem}.}   Investigations in this limit, supplemented if possible
by  an expansion in the inverse of the large parameter, often provide
valuable insights into the fundamental properties of a system even  when the
inverse parameter is not very small.

One of the best-known mean-field theories in many-body physics is the Weiss
molecular-field theory for the Ising model~\cite{Vollhardt:Baxter}. It is a
prototypical \emph{single-site mean-field theory} which becomes exact  for
infinite-range interaction, as well as in the limit of the  coordination
number $Z \to \infty$ or the dimension $d \to \infty$.  In the latter case $1/Z$ or $1/d$ is a small parameter which can  be used to improve the
mean-field theory systematically. This  mean-field theory is comprehensive in the sense discussed above. Namely, it contains no
unphysical singularities, is  applicable for all values of the input
parameters, i.e., the coupling  parameter, magnetic field, and temperature, and
is  diagrammatically controlled \cite{Vollhardt:Itzykson89}.

Itinerant quantum mechanical models such as the Hubbard model and its
generalizations are much more complicated than classical, Ising-type models.
Generally there do not even exist semiclassical approximations for such
models that might serve as a starting point for further investigations.
Under such circumstances the construction of a mean-field theory with the
comprehensive properties of the Weiss molecular field theory for the Ising
model will necessarily be much more complicated, too. Here the limit of high
spatial dimensions\index{limit of high dimensions} $d$ or coordination number $Z$ has again been extremely
useful since it provides the basis for the construction of a comprehensive
\emph{dynamical} mean-field theory (DMFT) for lattice fermions.

In this article we will first discuss (sec. \ref{Vollhardt:Lattice Fermions in the Limit of High Dimensions}) the limit of high
spatial dimensions $d$ for lattice fermions, the scaling of the hopping
amplitude which is necessary to obtain a meaningful limit $d \to \infty$ as
well as the simplifications of the many-body perturbation theory occurring
in this limit. In sec. \ref{Vollhardt:Dynamical Mean-Field Theory for Correlated
Lattice Fermions} the self-consistency equations obtained in the limit $d
\to \infty$ are derived which provide the basis for the
DMFT for correlated lattice fermions. An example for the many insights
gained by the DMFT is the Mott-Hubbard metal-insulator transition discussed
in sec. \ref{Vollhardt:The Mott-Hubbard Metal-Insulator Transition}. The application
of the DMFT to
real correlated materials is described in sec. \ref{Vollhardt:Theory of Electronic
Correlations in Materials}.
Brief introductions to the DMFT for correlated systems in the presence of
disorder (sec. \ref{Vollhardt:Electronic Correlations and Disorder}), correlated
bosons in optical lattices (sec. \ref{Vollhardt:DMFT for Correlated Bosons in Optical
Lattices}), and systems in non-equilibrium (sec. \ref{Vollhardt:DMFT for
Nonequilibrium}) are also given. In sec. \ref{Vollhardt:Summary and Outlook} a
summary and outlook is presented.

\section{Lattice Fermions in the Limit of High Dimensions}\index{limit of high dimensions}

\label{Vollhardt:Lattice Fermions in the Limit of High Dimensions}

\subsection{Scaling of the Hopping Amplitude}

We consider the kinetic energy term (\ref{Vollhardt:G11.7a}) since the
interaction term is purely local and is thereby completely independent of
the lattice structure and the dimension. For hopping between
nearest-neighbor (NN) sites $i$ and $j$ with amplitude $t_{ij}\equiv -t$ on
a $d$-di\-men\-sion\-al hypercubic  lattice with lattice spacing $a$, the dispersion $\epsilon_{%
\bm{k}}$\index{dispersion relation} is given by
\end{subequations}
\begin{equation}
\epsilon_{\bm{k}} = - 2t \sum_{n = 1}^{d} \; \cos (k_{n}a).  \label{Vollhardt:G11.8}
\end{equation}
The density of states (DOS)\index{density of states} corresponding to $\epsilon_{\bm{k}}$ is
\begin{equation}
N_d (\omega) = \sum_{\bm{k}} \delta (\hbar\omega - \epsilon_{\bm{k}}),
\label{Vollhardt:G11.9}
\end{equation}
which is the probability density for finding\footnote{%
In the following we set the Planck constant $\hbar $, the Boltzmann constant
$k_{B}$, and the lattice constant $a$ equal to unity.} $\omega = \epsilon_{\bm{k}}$  for a random choice of $\bm{k} = (k_1, \ldots, k_d)$. If the $k_i$
are  chosen randomly, $\epsilon_{\bm{k}}$ in \eqref{Vollhardt:G11.8} is the sum of
(independent) random numbers $-2t \cos k_i$. The central limit theorem then
implies \cite{Vollhardt:metzner89} that in the limit $d \to \infty$ the DOS is given by a Gaussian
\begin{equation}
N_d (\omega) \overset{d \to \infty}{\longrightarrow} \; \frac{1}{2t \sqrt{\pi d}} \exp \Bigg[ - \Big( \frac{\omega}{2t \sqrt{d}} \Big)^2 \Bigg].
\label{Vollhardt:G11.10}
\end{equation}
Unless $t$ is scaled properly with $d$ this DOS will become arbitrarily
broad and featureless for  \hbox{$d \to \infty$}. Clearly only the
scaling\index{scaling of hopping amplitudes}
\begin{equation}
t \to \frac{t^*}{\sqrt{d}}, \; t^* = \mathrm{const.},  \label{Vollhardt:G11.11}
\end{equation}
(``quantum scaling'') yields a non-trivial limit $d \to \infty$ for the DOS \cite{Vollhardt:Wolff83,Vollhardt:metzner89}.

The interaction term in \eqref{Vollhardt:G11.7} is seen to be purely local and
independent  of the surrounding; hence it is independent of the spatial
dimension of the system.  Consequently, the on-site interaction $U$ need not
be scaled.  So we see that the scaled Hubbard Hamiltonian
\begin{equation}
\hat{H} = - \frac{t^*}{\sqrt{Z}} \; \sum_{\langle i, j \rangle}
\sum\limits_\sigma \; \hat{c}_{i \sigma}^{+} \hat{c}_{j\sigma} + U \sum_{i}
\hat{n}_{i \uparrow} \hat{n}_{i \downarrow}  \label{Vollhardt:G11.18}
\end{equation}
has a nontrivial $Z \to \infty$ limit, where
both terms, the kinetic energy and the interaction, are of  the same order
of magnitude and are thereby able to compete; here $\langle i, j \rangle$
denotes NN sites $i$ and $j$. It is this competition between the two terms
which leads to interesting many-body physics. Mathematically this is
expressed by the fact that the generic matrix elements of the commutator between the kinetic and the
interaction part of the Hamiltonian do not vanish in the $d\to \infty $
limit.

The quantum scaling \eqref{Vollhardt:G11.11} was determined within a $\bm{k}$-space
formulation, but it can also be derived within a position-space formulation
as will be discussed next.

\subsection{Simplifications of the Many-Body Perturbation Theory}

\label{Vollhardt:sec:scaling}

The most important consequence of the scaling \eqref{Vollhardt:G11.11} is the fact
that it leads to significant simplifications in the investigation  of
Hubbard-type lattice models \cite{Vollhardt:metzner89,Vollhardt:MH89a,Vollhardt:metzner89a,Vollhardt:MH89b,Vollhardt:Czycholl__1of4,Vollhardt:Czycholl__4of4,Vollhardt:Brandt}. To understand this point better we take a look at the perturbation theory
in terms of $U$.  At $T = 0 $ and $U = 0$ the kinetic energy of the electrons
is given by
\begin{equation}
E_{\mathrm{kin}}^0 = - t \sum_{\langle i, j \rangle} \sum_{\sigma} \; g_{ij,
\sigma}^{0}.  \label{Vollhardt:G11.19}
\end{equation}
Here $g_{ij , \sigma}^0 = \langle \hat{c}_{i \sigma}^{+} \hat{c}_{j
\sigma}^{} \rangle_0$  is the one-particle density matrix which can
be interpreted as the probability amplitude for hopping from site $j$ to site $i$. The  square of its absolute value is proportional to the probability for an electron to hop from $j$ to $i$, i.e., $| g_{ij ,
\sigma}^0 |^2 \sim 1/Z \sim 1/d$, since site $j$ has $\mathcal{O}(d)$ NN sites $i$. The sum of $| g_{ij , \sigma}^0 |^2$  over all NN sites $i$ of $j$ must then yield a $Z$ or $d$ independent constant.  In the limit $d \to \infty$ we therefore find
\begin{equation}
g_{ij, \sigma}^0 \sim \mathcal{O} \Big( \frac{1}{\sqrt{d}} \Big).  \label{Vollhardt:G11.20}
\end{equation}
Since the sum over NN sites in \eqref{Vollhardt:G11.19} is of  $\mathcal{O}(d)$,
the NN hopping amplitude $t$ must obviously be scaled  according to
\eqref{Vollhardt:G11.11} for $E_{\mathrm{kin}}^0$ to remain finite in  the limit $d, Z
\to \infty$. Hence, as expected, a real-space  formulation yields the same
results for the required scaling of the  hopping amplitude.

The one-particle Green function\index{Green function}  $G_{ij , \sigma}^0 (\omega)$ of the
non-interacting system  obeys the same scaling as $g_{ij , \sigma}^0$.  This
follows directly from its definition
\begin{equation}
G_{ij , \sigma}^0(t)\equiv -\langle T \hat{c}_{i\sigma}(t) \hat{c}^+_{j\sigma}(0) \rangle_0,  \label{Vollhardt:G11.21}
\end{equation}
where $T$ is the time ordering operator, and the time evolution of the
operators is given by the Heisenberg representation. The one-particle
density matrix is obtained as $g_{ij,\sigma}^0 =\lim_{t\rightarrow 0^-}
G^0_{ij,\sigma}(t)$. If $g_{ij,\sigma}^0$ obeys (\ref{Vollhardt:G11.20}) the
one-particle Green function must follow the same scaling at all times since
this property does not dependent on the time evolution and the quantum
mechanical representation. The Fourier transform $G_{ij , \sigma}^0 (\omega)$
also preserves this property.

It is important to realize that, although $G_{ij , \sigma}^0 \sim 1/\sqrt{d}$
vanishes  for $d \to \infty$, the particles are \emph{not} localized, but
are still mobile. Indeed,  even in the limit  $d \to \infty$ the
off-diagonal elements of $G_{ij , \sigma}^0$ contribute, since a particle
may  hop to $d$ nearest neighbors with reduced amplitude $t^*/\sqrt{2d}$.
For general $i, j$ one finds  \cite{Vollhardt:vanDongen89,Vollhardt:metzner89a}
\begin{equation}
G_{ij , \sigma}^0 \sim \mathcal{O} \Big( 1/d^{|| \bm{R}_i - \bm{R}_j
||/2 } \Big),  \label{Vollhardt:G11.22}
\end{equation}
where $|| \bm{R} || = \sum_{n = 1}^{d} |R_n|$ is the
length of $\bm{R}$ in  the so-called ``New York metric'' (also called ``taxi
cab metric'', since particles  only hop along horizontal or vertical lines,
never along a diagonal).

It is the property \eqref{Vollhardt:G11.22}
which  is the origin of all simplifications arising in the limit  $d \to
\infty$. In particular, it implies the collapse of all  connected,
irreducible perturbation theory diagrams\index{collapse of diagrams} in position space  \cite{Vollhardt:metzner89,Vollhardt:MH89a,Vollhardt:metzner89a}. In general, any two vertices  which are
connected by more than two separate paths\footnote{Here a ``path''  is any sequence of lines in a diagram; they are
``separate'' when they have no lines  in common.}  will collapse onto the same site.
 In particular, the
external vertices of any irreducible self-energy diagram are  always
connected by three separate paths and hence always  collapse. As a
consequence the full irreducible self-energy\index{self-energy!locality of} becomes a purely local
quantity \cite{Vollhardt:metzner89}, but retains its dynamics  \cite{Vollhardt:MH89a}
\begin{subequations}
\label{Vollhardt:G11.23}
\begin{equation}
\Sigma_{ij, \sigma} (\omega) \overset{d \to \infty}{=} \Sigma_{ii,
\sigma}(\omega) \delta_{ij}.  \label{Vollhardt:G11.23a}
\end{equation}
In the paramagnetic phase we may write $\Sigma_{ii, \sigma}(\omega) \equiv
\Sigma (\omega)$.  The Fourier transform of $\Sigma_{ij , \sigma}$ is then  momentum-independent
\begin{equation}
\Sigma_{\sigma} (\bm{k}, \omega) \overset{d \to \infty}{\equiv}
\Sigma_{\sigma} (\omega).  \label{Vollhardt:G11.23b}
\end{equation}
This leads to tremendous simplifications in all many-body  calculations for
the Hubbard model and related models. It should be  noted that a $\bm{k}$-independence of $\Sigma$ was sometimes  \emph{assumed} as a convenient
approximation (``local  approximation'') \cite{Vollhardt:Kajzar78,Vollhardt:Treglia80,Vollhardt:Bulk90}.
Here we  identified the limit where this is indeed exact.


\subsection{Interactions Beyond the On-Site Interaction}

\label{Vollhardt:sec:selfenergy}

In the case of more general interactions than the Hubbard interaction,
e.g., nearest-neighbor interactions such as
\end{subequations}
\begin{equation}
\hat{H}_{nn} = \sum_{\langle i , j \rangle} \; \sum_{\sigma \sigma^{\prime}}
\; V_{\sigma \sigma^{\prime}} \hat{n}_{i \sigma}^{} \hat{n}_{j
\sigma^{\prime}}^{}  \label{Vollhardt:G11.30}
\end{equation}
the interaction constant has to be scaled, too, in the limit  $d \to \infty$. In the case of \eqref{Vollhardt:G11.30}, which has the form  of a classical
interaction, the ``classical'' scaling
\begin{equation}
V_{\sigma \sigma^{\prime}} \to \frac{V_{\sigma \sigma^{\prime}}^*}{Z}
\label{Vollhardt:G11.31}
\end{equation}
is required. Of course, the propagator  still has the dependence
\eqref{Vollhardt:G11.22}.

Due to \eqref{Vollhardt:G11.31} all contributions, except for the Hartree term, are
found to vanish in $d = \infty$  \cite{Vollhardt:MH89a}. Hence nonlocal interactions only contribute
via their Hartree contribution, which is purely static. This gives  the
Hubbard interaction a unique role: of all interactions for fermionic lattice
models only the purely local Hubbard interaction remains dynamical in the limit $d \to
\infty$  \cite{Vollhardt:MH89a}.

\subsection{Single-Particle Propagator}

Due to the $\bm{k}$-independence of the irreducible self-energy\index{self-energy},
\eqref{Vollhardt:G11.23b}, the  one-particle propagator of an interacting lattice
fermion system is given by
\begin{equation}
G_{\bm{k}, \sigma}^{} (\omega) = \frac{1}{\omega - \epsilon_{\bm{k}} + \mu -
\Sigma_{\sigma} (\omega)}.  \label{Vollhardt:G11.32}
\end{equation}
Most importantly, the $\bm{k}$ dependence of $G_{\bm{k}} (\omega)$ comes
entirely from the energy dispersion  $\epsilon_{\bm{k}}$ of the \emph{non}-interacting  particles. This means that for a homogeneous system with  the
propagator
\begin{equation}
G_{ij , \sigma} (\omega) = L^{-1} \; \sum_{\bm{k}} G_{\bm{k}, \sigma}
(\omega) e^{i \bm{k} \cdot (\bm{R}_i - \bm{R}_j) }  \label{Vollhardt:G11.33}
\end{equation}
its local part, $G_{ii , \sigma}$, has the form \cite{Vollhardt:MH89b}
\begin{subequations}
\label{Vollhardt:G11.34}
\begin{eqnarray}
G_{ii , \sigma} (\omega) & = & L^{-1} \sum\limits_{\bm{k}} G_{\bm{k},
\sigma} (\omega) = \int\limits_{-\infty}^{\infty} d E \frac{N_{\infty} (E)}{
\omega - E + \mu - \Sigma_{\sigma} (\omega)}   \label{Vollhardt:G11.34a} \\[12pt]
& \equiv & G_{\sigma}(\omega).  \label{Vollhardt:G11.34b}
\end{eqnarray}
In the following we will limit our discussion to the  paramagnetic phase and omit the spin index.
The spectral function\index{spectral function} of the  interacting system (often referred to as the
``density of states''\index{density of states} (DOS) as in the  non-interacting case) is given by
\begin{equation}
A(\omega) = - \frac{1}{\pi} \mathrm{Im} G(\omega + i0^+);  \label{Vollhardt:G11.34c}
\end{equation}
for $U=0$ one has $A(\omega)\equiv N(\omega)$.  In the  limit $d \rightarrow
\infty$  two quantities then play the most  important role: the local
propagator $G(\omega)$ and the self-energy  $\Sigma (\omega)$.

\subsection{Consequences of the Momentum Independence of the Self-Energy}

We now discuss some more consequences of the $\bm{k}$-independence  of the
self-energy
as derived by M\"{u}ller-Hartmann  \cite{Vollhardt:MH89b}. Let us consider the Hubbard
model, or any one of its  generalizations, in the
paramagnetic phase, i.e., without a  broken symmetry. At $T = 0$ the one-particle Green function\index{Green function} \eqref{Vollhardt:G11.32} then reads
\end{subequations}
\begin{equation}
G_{\bm{k}} (\omega) = \frac{1}{\omega - \epsilon_{\bm{k}} + E_F - \Sigma
(\omega)}.  \label{Vollhardt:G11.39}
\end{equation}
In general, even when $\Sigma(\omega)$ is $\bm{k}$-dependent, the Fermi
surface  is defined by the $\omega = 0$ limit of the denominator of
\eqref{Vollhardt:G11.39} as
\begin{subequations}
\label{Vollhardt:G11.40}
\begin{equation}
\epsilon_{\bm{k}} + \Sigma_{\bm{k}} (0) = E_F.  \label{Vollhardt:G11.40a}
\end{equation}
According to Luttinger and Ward \cite{Vollhardt:LW60} the volume within the  Fermi
surface is not changed by interactions, provided the effect  of the latter
can be treated in infinite-order perturbation theory  (i.e., no broken
symmetry). This is expressed by
\begin{equation}
n = \sum_{\bm{k} \sigma} \; \theta [E_F - \epsilon_{\bm{k}} - \Sigma_{\bm{k}%
} (0) ],  \label{Vollhardt:G11.40b}
\end{equation}
where $n$ is the particle density and $\theta (x)$ is the step  function. In
general, the $\bm{k}$-dependence of $\Sigma_{\bm{k}}  (0)$ in \eqref{Vollhardt:G11.40a}
implies that, in spite of \eqref{Vollhardt:G11.40b},  the shape of the Fermi surface
of the interacting system will be quite  different from that of the
non-interacting system (except for the  fully rotation invariant case $\epsilon_{\bm{k}} \sim k^2)$. For  lattice fermion models in $d < \infty$,
with $\Sigma_{\bm{k}}  (\omega) \equiv \Sigma (\omega)$, \eqref{Vollhardt:G11.40a}
implies that the  Fermi surface itself (and hence the volume enclosed) is
not  changed by interactions.\footnote{In $d=\infty$ limit the notion of a Fermi surface of a lattice system is complicated by the fact that the dispersion $\epsilon_{\bm{k}}$ is not a simple smooth
function.} The Fermi energy is simply shifted  uniformly from its
non-interacting value $E_F^0$, i.e., $E_F =  E_F^0 + \Sigma (0)$, to keep $n
$ in \eqref{Vollhardt:G11.40b} constant.  From \eqref{Vollhardt:G11.34a} we thus conclude that
the $\omega = 0$ value  of the local propagator, $G(0)$, and hence of the
spectral  function, $A(0) = - \frac{1}{\pi} \mathrm{Im} G(i0^+)$, is not
changed by interactions.
Renormalizations of $A(0)$ can only come from a  $\bm{k}$-dependence of $%
\Sigma$, i.~e., if $\partial \Sigma  /\partial \bm{k} \neq 0$.

For $\omega \to 0 $ the self-energy has the property
\begin{equation}
\mathrm{Im} \; \Sigma (\omega) \propto \omega^2  \label{Vollhardt:G11.40c}
\end{equation}
which implies quasiparticle (Fermi liquid)\index{Fermi liquid} behavior. The effective mass
\begin{equation}
\frac{m^*}{m} = \left. 1 - \frac{d \Sigma}{d \omega} \right|_{\omega = 0} =
1 + \frac{1}{\pi} \; \int_{-\infty}^{\infty} \; d \omega \; \frac{\mathrm{Im}
\Sigma (\omega + i0^-)}{\omega^2} \geq 1  \label{Vollhardt:G11.40d}
\end{equation}
is seen to be enhanced. In particular, the momentum distribution
\end{subequations}
\begin{equation}
n_{\bm{k}} = \frac{1}{\pi} \; \int_{-\infty}^{0} \; d \omega \; \mathrm{Im}
G_{\bm{k}} (\omega)  \label{Vollhardt:G11.41}
\end{equation}
has a discontinuity at the Fermi surface\index{Fermi surface discontinuity}, given by  $n_{k_{F}^-} - n_{k_F^+}
= (m^*/m )^{-1}$, where $k_F^\pm = k_F \pm 0^+$.

\section{Dynamical Mean-Field Theory for Correlated Lattice Fermions}

\label{Vollhardt:Dynamical Mean-Field Theory for Correlated Lattice Fermions}

The limit of high spatial dimensions $d$ or coordination number  $Z$
provides  the basis for the construction of a comprehensive mean-field
theory  for lattice fermions which is diagrammatically controlled and whose
free energy has no unphysical singularities.  It starts from
the scaled Hamiltonian (\ref{Vollhardt:G11.18}) and makes use of the simplifications in the
many-body perturbation theory discussed in sec. \ref{Vollhardt:sec:scaling}. There we
found that the local propagator $G(\omega)$, i.e., the probability amplitude for an
electron to return to a lattice site, and the local, but fully dynamical
self-energy $\Sigma (\omega)$ are the most important quantities in  such a
theory. Since the self-energy is a dynamical variable (in  contrast to the
Hartree-Fock theory\index{Hartree-Fock method} where it is merely an average, static  potential) the
resulting mean-field theory is also dynamical  and can thus describe genuine
correlation effects such as the  Mott-Hubbard metal-insulator transition.

The self-consistency equations of this \emph{dynamical mean-field theory}\index{dynamical mean-field theory}
(DMFT) for correlated lattice fermions can be derived in  different ways.
All derivations make use of the fact  that in the limit of high spatial
dimensions Hub\-bard-type models, i.e., lattice models with a local interaction, reduce to a ``dynamical single-site
problem'',\index{dynamical single-site problem} where the  $d$-di\-men\-sion\-al lattice model is effectively described
by the  dynamics of the correlated fermions on a single site  embedded in a
``bath'' provided by the other particles.
\begin{figure}[tbp]
\centerline{\includegraphics[width=0.8\textwidth]{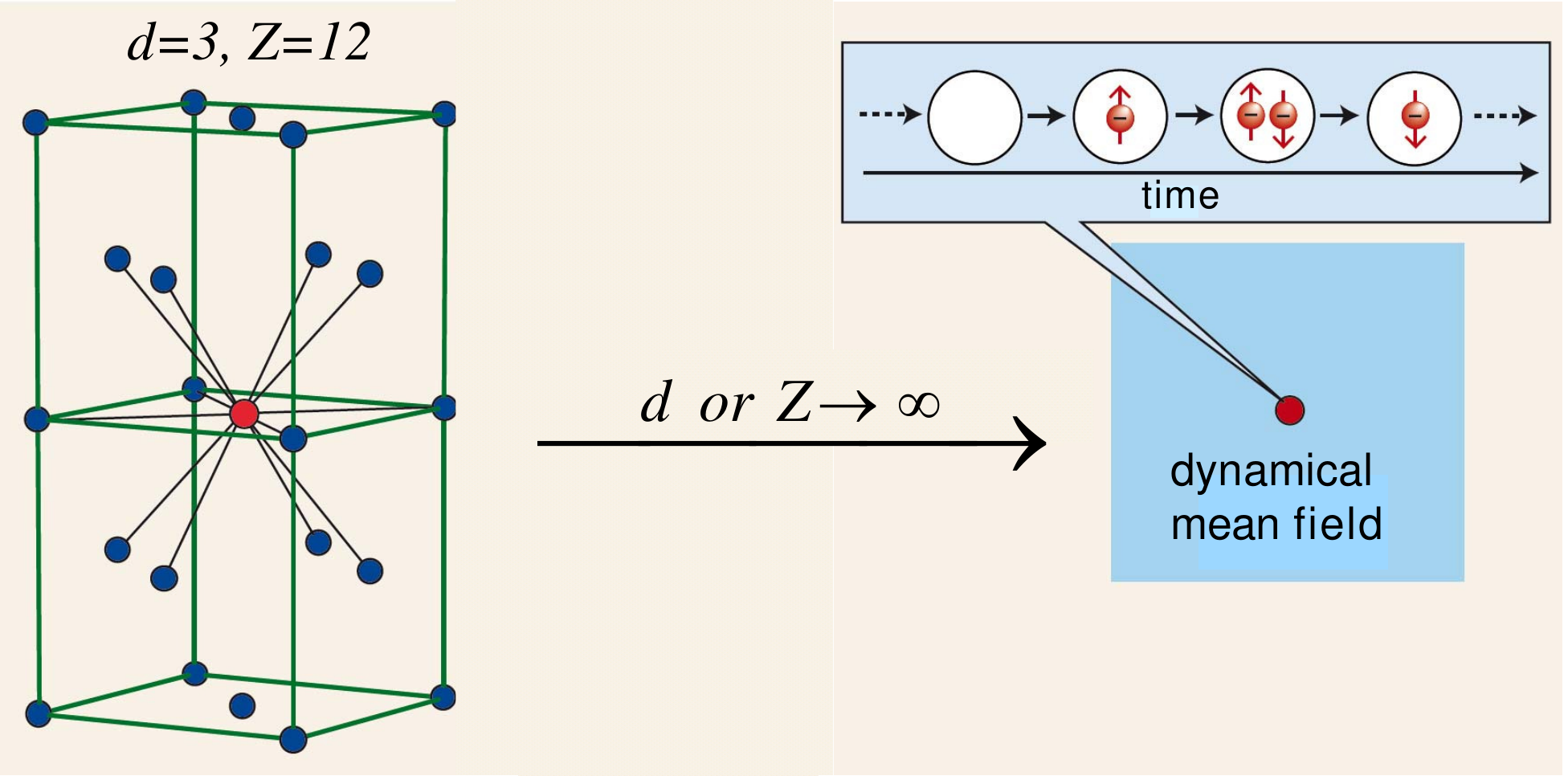}}
\caption[DMFT and the limit of high dimensions and coordination numbers.]{Already in a $d=3$ the coordination number $Z$ can be quite large,
as in the case of a face-centered cubic lattice with $Z=12$. In the limit $d\rightarrow \infty $, i.e., $Z\rightarrow \infty $, the many-body lattice
problem reduces to that of a single lattice site embedded in a dynamical mean
field. As shown in the inset electrons can hop onto and off that site and interact
as in the finite-dimensional Hubbard model. Therefore the DMFT describes the dynamics of the
interacting electrons correctly.}
\label{Vollhardt:DMFT}
\end{figure}
In particular, the derivation by Jani\v{s}  \cite{Vollhardt:Janis91,Vollhardt:Janis92b} is a
generalization of the coherent potential approximation\index{coherent potential approximation} (CPA)\index{CPA|see{coherent potential approximation}} for disordered
systems\footnote{The CPA is the best single-site approximation for disordered, non-interacting lattice electrons \cite{Vollhardt:Velicky68,Vollhardt:Yonezawa73,Vollhardt:Elliot74}; it becomes exact in the limit $d,Z \rightarrow \infty$ \cite{Vollhardt:vlaming92,Vollhardt:Vollhardt-Salerno}.} to the Hubbard model.   In the following we will present  today's
standard derivation by Georges and  Kotliar \cite{Vollhardt:Georges92} which is based
on the mapping of the  lattice problem onto a self-consistent
single-impurity Anderson  model; this approach was also employed by Jarrell
\cite{Vollhardt:Jarrell92}.
Although the DMFT equations derived within the CPA
    approach and the self-consistent single-impurity approach, respectively, are
    identical it is the latter formulation which
was immediately adopted by the community since it makes  contact with the
theory of quantum impurities and Kondo problems;\index{Kondo problem} for a review see \cite{Vollhardt:georges96}. This is a well-understood  branch of many-body physics \cite{Vollhardt:Fabrizio} for whose
solution efficient numerical  codes had been developed already in the
1980's, in particular by use of the quantum Monte-Carlo\index{quantum Monte Carlo method} (QMC)\index{QMC|see{quantum Monte Carlo method}} method
\cite{Vollhardt:Hirsch86}.  \newline
\newline

\subsection{Construction of the DMFT as a Self-Consistent Single-Impurity
Anderson Model}

\label{Vollhardt:sec:cavity}

Following the presentation of Georges, Kotliar, Krauth, and Rozenberg \cite{Vollhardt:georges96} the
DMFT equations will now be  derived using the so-called cavity method\footnote{We note that the sign of the hopping amplitude $t_{ij}$ used here (see the definition in \eqref{Vollhardt:G11.7b}) is opposite to that in ref. \cite{Vollhardt:georges96}.}.\index{cavity method} This
derivation  starts by removing one lattice site together with its bonds from
the  rest of the lattice. The remaining lattice, which now contains a
cavity, is replaced by a particle bath which plays the role of the
dynamical mean field (see Fig.~\ref{Vollhardt:DMFT}). Now comes a
physically motivated step: the bath is coupled, via a hybridization, to the
cavity. The resulting problem then  amounts to the solution of an effective
single-impurity Anderson model\index{single-impurity Anderson model} where the degrees of freedom of the bath,
represented by an  appropriate hybridization function, have to be determined
self-consistently.

To be specific, we consider the partition function in the grand
canonical ensemble
\begin{equation}
{\mathcal{Z}} = \int \prod_{i \, \sigma} D c^{*} _{i \sigma} D c_{i \sigma}
\exp[-S].
\end{equation}
The action $S$ for the Hubbard model is  given by
\begin{multline}
S=\int_0^{\beta} d \tau \left[ \sum_{i\sigma} c^{*}_{i \sigma} (\tau) (\frac{\partial}{\partial \tau} - \mu) c_{i \sigma} (\tau) + \sum_{ij \,\sigma}
t_{ij} c^{*}_{i \sigma}(\tau) c_{j \sigma} (\tau) \right.  \\
\left. + \;\sum_{i}U
c^{*}_{i\uparrow}(\tau)c_{i\uparrow}(\tau)c^{*}_{i\downarrow}(\tau)c_{i\downarrow}(\tau) \right ],
\end{multline}
where
we use Grassmann variables  $c^{*} _{i \sigma}$, $c _{i \sigma}$. We split
the action $S$ into  three parts
\begin{equation}
S=S_{0}+\Delta S + S^{(0)},
\end{equation}
where $S_{0}$ is the part containing only variables on site $0$
\begin{equation}
S_0=\int_0^{\beta} d \tau \left[ \sum_{\sigma} c^{*}_{0 \sigma} (\tau) (\frac{\partial}{\partial \tau} - \mu) c_{0 \sigma} (\tau) + U
c^{*}_{0\uparrow}(\tau)c_{0\uparrow}(\tau)c^{*}_{0\downarrow}(\tau)c_{0\downarrow}(\tau) \right ],
\end{equation}
$\Delta S$ contains the hoppings between site $0$ and other sites of  the
lattice $i\neq 0$
\begin{equation}
\Delta S=\int_0^{\beta} d \tau \sum_{i \,\sigma} \left[ t_{i0} c^{*}_{i
\sigma}(\tau) c_{0 \sigma} (\tau) + t_{0i} c^{*}_{0 \sigma}(\tau) c_{i
\sigma} (\tau)\right ],
\end{equation}
and the rest, denoted by $S^{(0)}$, is the part of the  action where
the site $0$ and its bonds are removed, i.e., for $i,j\neq  0$ one has
\begin{multline}
S^{(0)}=\int_0^{\beta} d \tau \left[ \sum_{i\neq 0\,\sigma} c^{*}_{i \sigma}
(\tau) (\frac{\partial}{\partial \tau} - \mu) c_{i \sigma} (\tau) + \sum_{ij
\neq 0 \,\sigma} t_{ij} c^{*}_{i \sigma}(\tau) c_{j \sigma} (\tau) \right.
 \\
\left.+ \; U \sum_{i\neq 0}
c^{*}_{i\uparrow}(\tau)c_{i\uparrow}(\tau)c^{*}_{i\downarrow}(\tau)c_{i\downarrow}(\tau) \right ].
\end{multline}
We now rewrite the partition function ${\mathcal{Z}}$ as
\begin{multline}
{\mathcal{Z}} = \int \prod_{ \sigma }D c^{*}_{0\sigma} D c_{0\sigma} \exp[-S_{0}]  \\
\times\int \prod_{i\neq0\, \sigma } D c^{*}_{i\sigma} D c_{i\sigma} \exp[-S^{(0)}] \exp[-\Delta S]
\end{multline}
and use the  ensemble average
\begin{equation}
\langle X\rangle_{(0)} \equiv \frac{1}{ {\mathcal{Z}}^{(0)}} \int
\prod_{i\neq0\,\sigma } D c^{*}_{i\sigma} D c_{i\sigma} X\exp [-S^{(0)}]
\end{equation}
taken with respect to $S^{(0)}$ (the action where the site $i=0$ is
excluded), with ${\mathcal{Z}}^{(0)}$ being the corresponding  partition
function. Then the partition function reads
\begin{eqnarray}
{\mathcal{Z}} = {\mathcal{Z}}^{(0)}\int \prod_{\sigma }D c^{*}_{0\sigma} D
c_{0\sigma} \exp[-S_{0}] \langle \exp[-\Delta S]\rangle_{(0)}.
\end{eqnarray}
In the next step we expand the second exponent with respect to the  action $\Delta S$.
A non-trivial limit $d \rightarrow \infty$ is obtained by scaling  the
hopping amplitudes $t_{ij}$ as described in  sec.~\ref{Vollhardt:sec:scaling}.
Consequently, in the $Z\rightarrow \infty$  limit only the contribution $G_{jk\sigma}^{(0)}$, where
\begin{eqnarray}
G_{jk\sigma}^{(0)}(\tau_1-\tau_2)= - \langle T_{\tau}
c_{j\sigma}(\tau_1)c_{k\sigma}^*(\tau_2)\rangle_{{(0)}},
\end{eqnarray}
or disconnected contributions made of products of  $G_{jk\sigma}^{(0)}$'s
remain. Applying the linked-cluster theorem  and collecting only connected
contributions in the exponential  function one obtains the local action
\begin{multline}
S_{\mathrm{loc}}= \left[\int_0^{\beta} d \tau \sum_{\sigma} c^{*}_{0 \sigma}
(\tau) (\frac{\partial}{\partial \tau} - \mu) c_{0 \sigma} (\tau)\right.
\\
+ U\int_0^{\beta} d \tau
c^{*}_{0\uparrow}(\tau)c_{0\uparrow}(\tau)c^{*}_{0\downarrow}(\tau)c_{0%
\downarrow}(\tau)
\\
\left. + \int_0^{\beta}d\tau_1 \int_0^{\beta}d\tau_2 {\sum_{\sigma}}{%
\sum_{j,k\neq0}} t^{*}_{j0} t^{*}_{k0} G_{jk\sigma}^{(0)}(\tau_1-\tau_2)
c^*_{0\sigma}(\tau_1) c_{0\sigma}(\tau_2) \right],
\end{multline}
where the rescaled hoppings are denoted with a star. Introducing  the
hybridization function\index{hybridization function}
\begin{equation}
\Delta_{\sigma}(\tau_1-\tau_2)= - {\sum_{i,j\neq0}}t^{*}_{i0}t^{*}_{j0} {G}_{ij\sigma}^{(0)}(\tau_1-\tau_2),  \label{Vollhardt:hybridization_function}
\end{equation}
and employing the free (``Weiss'') mean-field propagator  $\mathcal{G}_{\sigma}$ one can express the DMFT local action in the  following form
(here the site index $i=0$ is omitted for simplicity)
\begin{multline}
S_{\mathrm{loc}}=- \int_0^{\beta} d \tau_1 \int_0^{\beta} d \tau_2
\sum_{\sigma} c^{*}_{\sigma} (\tau_1) \mathcal{G}^{-1}_{\sigma}(\tau_1-\tau_2) c _{\sigma}(\tau_2)  \\
+ \; U\int_0^{\beta} d \tau
c^{*}_{\uparrow}(\tau)c_{\uparrow}(\tau)c^{*}_{\downarrow}(\tau)c_{\downarrow}(\tau),  \label{Vollhardt:local_action}
\end{multline}
where
\begin{equation}
\mathcal{G}^{-1}_{\sigma}(\tau_1-\tau_2)=-\left( \frac{\partial}{\partial\tau_1}-\mu \right)
\delta_{\tau_1\tau_2}-\Delta_{\sigma}(\tau_1-\tau_2).
\label{Vollhardt:hybridization_function2}
\end{equation}

Finally, we need the relation between the Green function ${G}^{(0)
}_{ij\sigma}(\tau-\tau^{\prime})$ where the site $i=0$ is removed  and the
full lattice Green function, i.e.,
\begin{equation}
{G}^{(0) }_{ij\sigma}={G}_{ij\sigma}-{G}_{i0\sigma} {G}_{00\sigma}^{-1} {G}_{0j\sigma},  \label{Vollhardt:lattice_local}
\end{equation}
which holds for a general lattice.

In order to obtain the full solution of the lattice problem it is convenient
to express the relation between the local Green function ${G}_{00\sigma
}\equiv {G}_{\sigma }$ and the ``Weiss'' mean field\footnote{In principle, any one of the local functions $\mathcal{G}_{\sigma
}(i\omega _{n})$, $\Sigma _{\sigma }(i\omega _{n})$, or $\Delta
_{\sigma }(i\omega _{n})$ can be viewed as a ``dynamical
mean field''\protect\index{dynamical mean field} acting on particles on a site, since they all
appear in the bilinear term of the local action (\ref{Vollhardt:local_action}).} $\mathcal{G}_{\sigma }^{-1}$ in the form of a Dyson equation\index{Dyson equation}
\begin{eqnarray}
\lbrack G_{\sigma }(i\omega _{n})]^{-1} &=&[\mathcal{G}_{\sigma
}(i\omega _{n})]^{-1}-\Sigma _{\sigma }(i\omega _{n})  \label{Vollhardt:Bath_function} \\
&=&i\omega _{n}+\mu -\Delta _{\sigma }(i\omega _{n})-\Sigma _{\sigma
}(i\omega _{n}).  \label{Vollhardt:Dyson_eq}
\end{eqnarray}%
Then the lattice Green function (in ${\bm{k}}$-space) $G_{\bm{k}\,\sigma
}(i\omega _{n})$ is given by
\begin{equation}
G_{\bm{k}\,\sigma }(i\omega _{n})=\frac{1}{i\omega _{n}-\epsilon _{\bm{k}}+\mu -\Sigma _{\sigma }(i\omega _{n})}.
\end{equation}%
After performing the so-called lattice Hilbert transform we recover the
local Green function
\begin{eqnarray}
G_{\sigma }(i\omega _{n}) &=&\sum_{\bm{k}}G_{\bm{k}\,\sigma }(i\omega
_{n})=\sum_{\bm{k}}\frac{1}{i\omega _{n}-\epsilon _{\bm{k}}+\mu -\Sigma
_{\sigma }(i\omega _{n})} \\
&=&\underset{-\infty }{\overset{\infty }{\int }}d\epsilon \frac{N(\omega )}{i\omega _{n}-\epsilon +\mu -\Sigma _{\sigma }(i\omega _{n})}. \label{Vollhardt:local_prop}
\end{eqnarray}
The ionic lattice on which the electrons move, and its structure, are seen to enter only via the DOS of the non-interacting electrons.
After analytic continuation to real frequencies the local (\textquotedblleft
$\bm{k}$ averaged\textquotedblright ) propagator reads
\begin{equation}
G_{\bm{k}\,\sigma }(\omega )=\frac{1}{\omega -\epsilon _{\bm{k}}+\mu -\Sigma
_{\sigma }(\omega )}.
\end{equation}

This completes the derivation of the self-consistent DMFT equations\index{DMFT equations}. Namely, the functional integral determining the local propagator
\begin{equation}
G_{\sigma }(i\omega _{n})=-\frac{1}{{\mathcal{Z}}}\int \prod_{\sigma
}Dc_{\sigma }^{\ast }Dc_{\sigma }[c_{\sigma }(i\omega _{n})c_{\sigma }^{\ast
}(i\omega _{n})]\exp [-S_{\mathrm{loc}}],
\label{Vollhardt:impurity_problem}
\end{equation}
where the partition function $\mathcal{Z}$ and the local action $S_{\mathrm{loc}}$ are given by
\begin{equation}
{\mathcal{Z}}=\int \prod_{\sigma }Dc_{\sigma }^{\ast }Dc_{\sigma }\exp [-S_{\mathrm{loc}}]
\end{equation}
and \eqref{Vollhardt:local_action}, respectively, together with the expression \eqref{Vollhardt:local_prop} for the lattice Green function provide a closed set of
equations for the local propagator $G_{\sigma }(i\omega _{n})$ and the self-energy $\Sigma
_{\sigma }(i\omega _{n})$. These equations
can be solved iteratively: Starting with an initial value for the self-energy one obtains the local propagator from \eqref{Vollhardt:local_prop} and thereby the bath Green function $\mathcal{G}_{\sigma
}(i\omega _{n})$ from \eqref{Vollhardt:Bath_function}. This determines the local action \eqref{Vollhardt:local_action} which is used to solve the single-impurity problem \eqref{Vollhardt:impurity_problem}, leading to a new value for the local propagator and, by employing the old self-energy, a new bath Green function, etc.
The
single-impurity problem is still a complicated many-body interacting problem
which cannot, in general, be solved exactly.

\subsection{Solution of the Self-Consistency Equations of the DMFT}

Due to the purely on-site interaction in the local action
\eqref{Vollhardt:local_action} the dynamics of the Hubbard model, \eqref{Vollhardt:G11.7},
remains complicated even in the limit $d \to \infty$.
Exact evaluations are only feasible when there is no  coupling between the
frequencies as, for example, in  the Falicov-Kimball\index{Falicov-Kimball model} model \cite{Vollhardt:fk_dmft}.
This model was solved  analytically by Brandt and Mielsch \cite{Vollhardt:Brandt} soon
after the  introduction of the $d \to \infty$ limit \cite{Vollhardt:metzner89}.

In general, the local action \eqref{Vollhardt:local_action} is the most complicated
part of the DMFT equations. To solve the self-consistency equations
different techniques (``impurity solvers'')\index{impurity solver} have been developed which are
either fully numerical and ``numerically exact'', or semi-analytic and
approximate.
The numerical solvers can be divided into renormalization group techniques
such as
the
numerical renormalization group\index{numerical renormalization group} (NRG)\index{NRG|see{numerical renormalization group}}
 \cite{Vollhardt:Bulla,Vollhardt:NRG-RMP} and the
density-matrix renormalization group\index{density-matrix renormalization group}
(DMRG)\index{DMRG|see{density-matrix renormalization group}}
\cite{Vollhardt:dmrg},
exact diagonalization\index{exact diagonalization}
(ED)\index{ED|see{exact diagonalization}}
\cite{Vollhardt:Caffarel94,Vollhardt:Si94,Vollhardt:Rozenberg94a}, and methods based
on the stochastic sampling of quantum and thermal averages, i.e.,
quantum Monte-Carlo\index{quantum Monte Carlo method}
(QMC) techniques such as the Hirsch-Fye QMC algorithm \cite{Vollhardt:Jarrell92,Vollhardt:Rozenberg92,Vollhardt:Georges92b,Vollhardt:georges96} and
continuous-time (CT)\index{CT-QMC|see{quantum Monte Carlo method continuous-time}} QMC\index{quantum Monte Carlo method!continuous-time}
\cite{Vollhardt:Rubtsov,Vollhardt:Werner,Vollhardt:Haule}.

Semi-analytic approximations such as the
iterated perturbation theory\index{iterated perturbation theory}
(IPT)\index{IPT|see{iterated perturbation theory}} \cite{Vollhardt:Georges92,Vollhardt:IPT,Vollhardt:georges96},
the
non-crossing approximation\index{non-crossing approximation}
(NCA)\index{NCA|see{non-crossing approximation}} \cite{Vollhardt:georges96,Vollhardt:psi-k},
the
fluctuation exchange approximation\index{fluctuation exchange approximation}
(FLEX)\index{FLEX|see{fluctuation exchange approximation}}
\cite{Vollhardt:Bickers89,Vollhardt:LichtKats99,Vollhardt:Kotliar06,Vollhardt:Drchal05},
the
local moment approach\index{local moment approach}
(LMA)\index{LMA|see{local moment approach}}
\cite{Vollhardt:Logan,Vollhardt:Kauch}, and the
parquet approximation\index{parquet approximation} \cite{Vollhardt:parquet}
can also provide valuable insight.

It quickly turned out that the DMFT is a powerful tool for the
investigation of electronic systems with strong correlations. It  provides a
non-perturbative and thermodynamically consistent  approximation scheme for
finite-di\-men\-sion\-al systems which is  particularly valuable for the study of
intermediate-coupling  problems where perturbative techniques fail \cite{Vollhardt:freericks95,Vollhardt:georges96,Vollhardt:Georges2003,Vollhardt:dmft_phys_today,Vollhardt:Vollhardt-Salerno}.

\section{The Mott-Hubbard Metal-Insulator Transition}

\label{Vollhardt:The Mott-Hubbard Metal-Insulator Transition}

The correlation induced transition between a paramagnetic metal and  a
paramagnetic insulator, referred to as ``Mott-Hubbard  metal-insulator
transition (MIT)''\index{metal-insulator transition}, is one of the most intriguing  phenomena in condensed
matter physics~\cite{Vollhardt:mott68,Vollhardt:mott90,Vollhardt:Gebhard}.  This transition is a
consequence of the competition between the  kinetic energy of the electrons
and their local interaction $U$.  Namely, the kinetic energy prefers the
electrons to move (a wave  effect) which leads to doubly occupied sites and
thereby to  interactions between the electrons (a particle effect). For
large  values of $U$ the doubly occupied sites become energetically very
costly. The system may reduce its total energy by localizing the  electrons.
Hence the Mott transition is a  localization-delocalization transition,
demonstrating the  particle-wave duality of electrons \cite{Vollhardt:dmft_phys_today}.

Mott-Hubbard MITs are found, for example, in transition metal oxides  with
partially filled bands. For such systems  band theory
typically predicts metallic behavior. The most famous  example is V$_{2}$O$_{3}$\index{V2O3@V$_{2}$O$_{3}$} doped with Cr~\cite{Vollhardt:McWhR,Vollhardt:McWhetal,Vollhardt:RMcWh}.  In particular,  below $T=380$\,K the  metal-insulator transition in paramagnetic (V$_{0.96}$Cr$_{0.04}$)$_{2}$O$_{3}$ is of
first order  \cite{Vollhardt:McWhetal}, with discontinuities in the
lattice parameters and  in the conductivity.
However, the two phases  remain isostructural.

Making use of the half-filled, single-band Hubbard model \eqref{Vollhardt:G11.7}
the Mott-Hubbard MIT was studied intensively in the past  \cite{Vollhardt:HubbardI,Vollhardt:RMcWh,Vollhardt:mott68,Vollhardt:mott90,Vollhardt:Gebhard}. Important early results  were
obtained by Hubbard~\cite{Vollhardt:Hub3} within a Green function  decoupling scheme,
and by Brinkman and Rice~\cite{Vollhardt:BR} within the  Gutzwiller variational method~\cite{Vollhardt:Gutzwiller}, both at $T=0$.  Hubbard's approach yields a continuous
splitting of the band into a  lower and upper Hubbard band, but cannot
describe quasiparticle  features. By contrast, the Gutzwiller-Brinkman-Rice
approach\index{Gutzwiller-Brinkman-Rice approach} gives a  good description of the low-energy, quasiparticle
behavior, but  cannot reproduce the upper and lower Hubbard bands. In the
latter  approach the MIT is signalled by the disappearance of the
quasiparticle peak.

To solve this problem the DMFT has been extremely valuable since it
provided detailed insights into the nature of the Mott-Hubbard MIT  for all
values of the interaction $U$ and temperature $T$  \cite{Vollhardt:georges96,Vollhardt:dmft_phys_today}.

\subsection{DMFT and the Three-Peak Structure of the Spectral Function}

\label{Vollhardt:subsec:threepeak}

The Mott-Hubbard MIT is monitored by the spectral function  $A(\omega)= -
\frac{1}{\pi} \mathrm{Im} G(\omega + i0^+)$ of the  correlated electrons\footnote{In the following we only consider the  paramagnetic phase, whereas magnetic
order is assumed to be  suppressed (``frustrated'').}
\cite{Vollhardt:georges96}.  While at small $U$ the system can be
described by coherent  quasiparticles whose DOS still resembles that of the
free electrons,  the spectrum in the Mott insulator state consists of two
separate  incoherent ``Hubbard bands''\index{Hubbard bands} whose centers are separated
approximately by the energy $U$. The latter originate from  atomic-like
excitations at the energies $\pm U/2$ broadened by the  hopping of electrons
away from the atom. At intermediate values of  $U$ the spectrum then has a
characteristic three-peak structure as  in the single-impurity Anderson
model, which includes both the  atomic features (i.e., Hubbard bands) and
the narrow quasiparticle peak\index{quasiparticle peak} at low excitation energies, near $\omega=0$.
This corresponds  to a strongly correlated metal. The general structure of
the spectrum (lower  Hubbard band, quasiparticle peak, upper Hubbard band)
is rather  insensitive to the specific form of the DOS of the
non-interacting  electrons.  The width of the quasiparticle peak vanishes
for $U\to U_{\mathrm{c2}}(T)$. 
On decreasing $U$, the transition from the  insulator to the metal occurs at
a lower critical value $U_{\mathrm{c1}}(T)$, where the gap vanishes.

It is important to note that the three-peak spectrum originates from  a
lattice model with only \emph{one} type of electrons. This is in  contrast
to the single-impurity Anderson model whose spectrum shows  very similar
features, but is due to \emph{two} types of electrons --- the localized
orbital at the impurity site and the free  conduction band. Therefore the
screening of the magnetic moment  which gives rise to the Kondo effect in
impurity systems has a somewhat different origin in interacting lattice systems. Namely, as
explained by the  DMFT the electrons provide both the local
moments and  the electrons which screen these moments \cite{Vollhardt:georges96}.

Interestingly, for any typical spectral function $A(\omega)$ with three  peaks,
Kramers-Kronig relations and the DMFT self-consistency equations imply that
the self-energy $\Sigma(\omega)$ abruptly  changes slope \emph{inside} the
central peak at some frequency $\omega_\star$  \cite{Vollhardt:byczuk07}, once at positive and once
at negative frequency.  While this behavior is not visible in $A(\omega)$ itself,
it leads  to ``kinks''\index{kinks in the dispersion relation} in the effective dispersion relation $E_{\bm{k}}$\index{dispersion relation} of  one-particle excitations, which is defined as the frequency for
which the momentum-resolved spectral function $A(\bm{k},\omega)$ $=$  $-\mathrm{Im}G(\bm{k},\omega)/\pi$ $=$ $-(1/\pi)\mathrm{\ Im}[1/(\omega+\mu-\epsilon_{ \mathbf{k}}-\Sigma(\omega))]$ is  maximal. For
frequencies below $\omega_\star$ the dispersion is  given by Fermi-liquid\index{Fermi liquid}
(FL) theory,  $E_{\bm{k}}=Z_{\text{FL}}\epsilon_{\bm{k}}$, where $Z_{\mathrm{FL}}$  $=$ $(\partial \mathrm{Re}\Sigma(\omega)/\partial\omega)_{\omega=0}$
is  the FL renormalization parameter. The FL regime terminates at the  kink
energy scale $\omega_\star$. This energy cannot be obtained within  FL theory
itself. Namely, it is determined by $Z_{\mathrm{FL}}$ and the
non-interacting DOS, e.g., $\omega_\star=0.41Z_{\mathrm{FL}}D$, where  $D$
is an energy scale of the non-interacting system such as half the  bandwidth
\cite{Vollhardt:byczuk07}. Above $\omega_\star$ the dispersion is  given by a
different renormalization with a small offset,  $E_{\bm{k}}$ $=$ $Z_{\text{CP}}\epsilon_{\bm{k}}+\text{const}$,  where $Z_{\text{CP}}$ is the weight of
the central peak of  $A(\omega)$. This theory explains kinks in the slope of
the  dispersion as a direct consequence of the electronic  interaction~\cite{Vollhardt:byczuk07}. The same mechanism may also lead to  kinks in the
low-temperature electronic specific  heat~\cite{Vollhardt:Toschi09}. These kinks have
also been linked to maxima in  the spin susceptibility~\cite{Vollhardt:Uhrig09}. Of
course, \emph{additional}  kinks in the electronic dispersion may also arise
from the coupling  of electrons to bosonic degrees of freedom, such as
phonons  \cite{Vollhardt:Lanzara2002,Vollhardt:Shen2002} or spin fluctuations  \cite{Vollhardt:He2001,Vollhardt:Hwang2004}. Interestingly, recent  experiments~\cite{Vollhardt:Nickel09} have
found evidence for kinks in  Ni(110), which may be due to the electronic
mechanism discussed  here.

\begin{figure}[t]
\centerline{\includegraphics[width=0.7\textwidth]{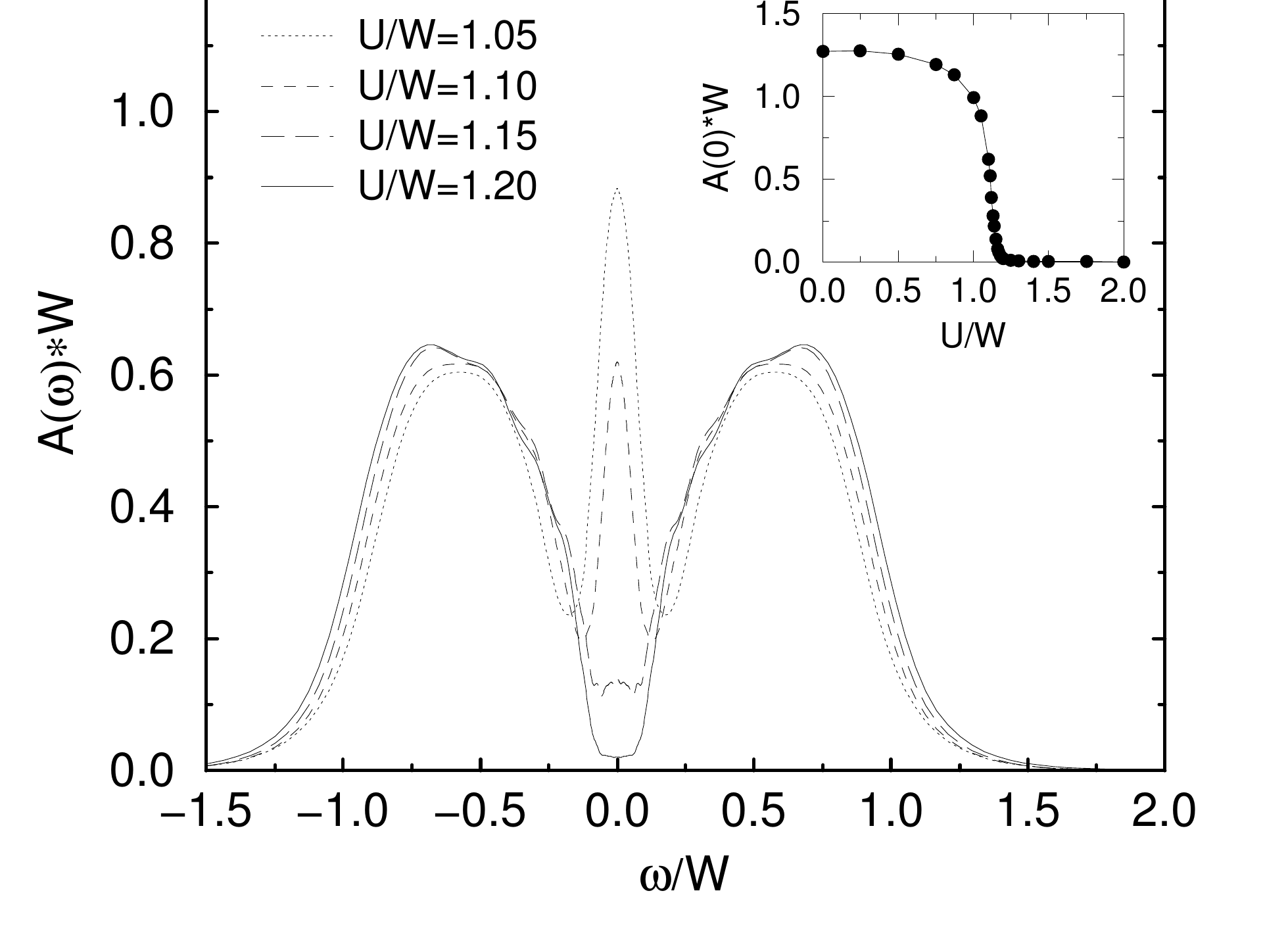}}
\caption[Spectral function of the half-filled Hubbard model.]{Spectral function for the half-filled Hubbard model for various
values of $U$ at $T=0.0276~W$ ($W$: bandwidth) in the crossover region. The
crossover from the metal to the insulator occurs via a gradual suppression
of the quasiparticle peak at $\protect\omega\!=\!0$. The inset shows the $U$
dependence of $A(\protect\omega\! =\!0)$, in particular the rapid decrease
for $U\approx 1.1~W$; from Ref.~\protect\cite{Vollhardt:Bulla01}.}
\label{Vollhardt:fig:4.1}
\end{figure}

The evolution of the spectral function of the half-filled frustrated
Hubbard model at a finite temperature is shown in  Fig.~\ref{Vollhardt:fig:4.1}.
This temperature is above the temperature of the  critical point so that
there is no real transition but only a  crossover from a metallic-like to an
insulating-like solution. The  height of the quasiparticle peak at the Fermi
energy is no longer  fixed at its zero temperature value. This is due to the temperature dependent imaginary part of the self-energy. The spectral weight of the
quasiparticle peak is seen to be gradually redistributed and shifted  to the
upper (lower) edge of the lower (upper) Hubbard band. The  inset of Fig.~\ref{Vollhardt:fig:4.1} shows the $U$-dependence of the spectral  function at zero
frequency $A(\omega\!=\!0)$. For higher values of  $U$ the spectral density
at the Fermi level is still finite and  vanishes only in the limit $U\to\infty$.
For
the insulating phase DMFT predicts the filling of the  Mott-Hubbard gap with
increasing temperature. This is due to the  fact that the insulator and the
metal are not distinct phases in  the crossover regime, implying that the
insulator has a finite  spectral weight at the Fermi level. This behavior
has  been detected experimentally by photoemission  experiments \cite{Vollhardt:Mo04}.


Altogether, the thermodynamic transition line $U_{\mathrm{c}}(T)$
corresponding to the Mott-Hubbard MIT\index{metal-insulator transition} is found to be of first order  at
finite temperatures, and is associated with a hysteresis region in  the
interaction range $U_{\mathrm{c1}}<U<U_{\mathrm{c2}}$, where $U_{\mathrm{c 1}}
$  and $U_{\mathrm{c 2}}$ are the interaction values at which the insulating and
metallic solution, respectively, vanish
\cite{Vollhardt:georges96,Vollhardt:Bulla,Vollhardt:Roz99,Vollhardt:Joo,Vollhardt:Bulla01,Vollhardt:Bluemer-Diss}.
\begin{figure}[tbp]
\centerline{\includegraphics[width=0.8\textwidth]{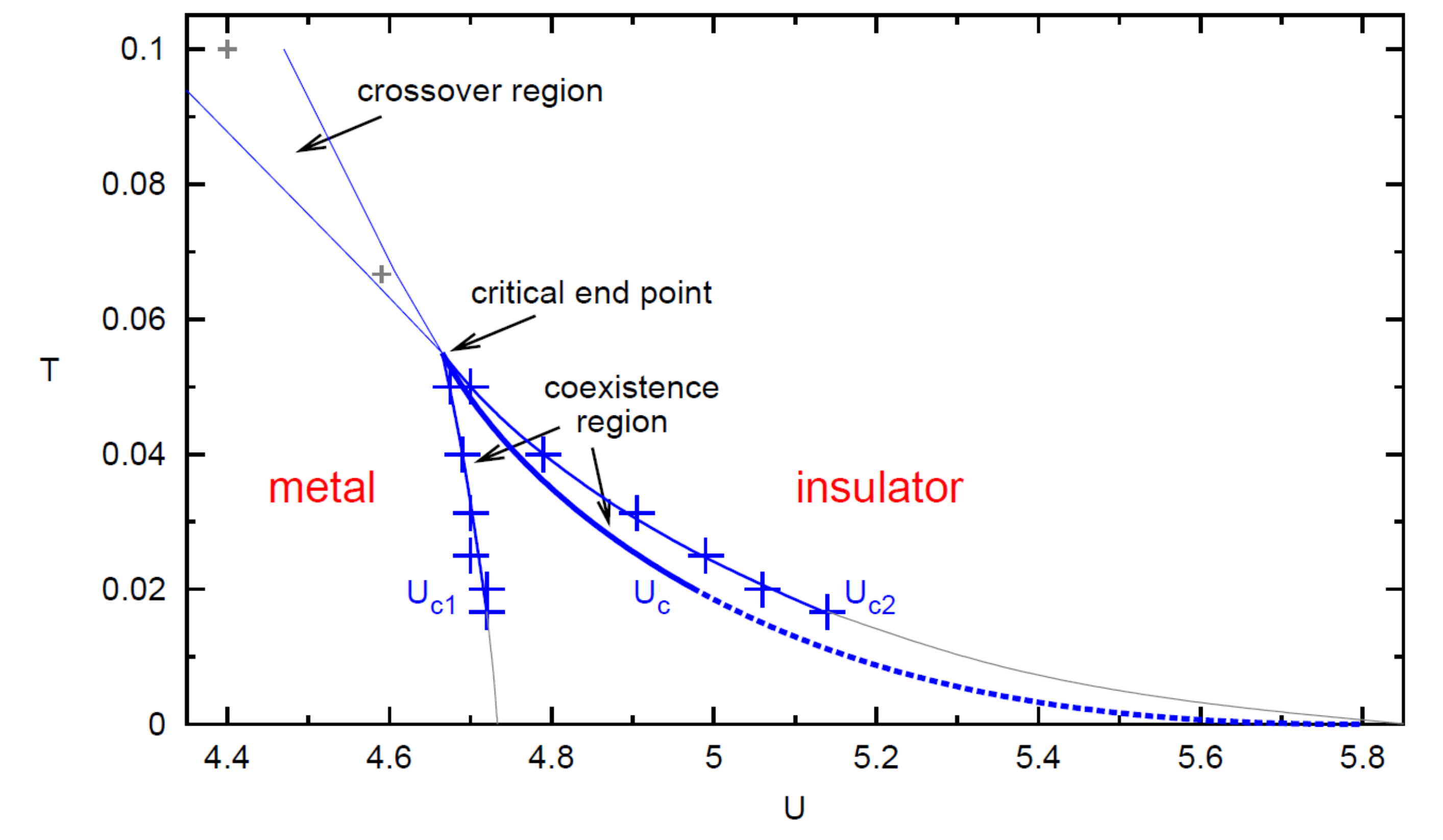}}
\caption[Phase diagram of the metal-insulator transition in the Hubbard model.]{Phase diagram of the Mott-Hubbard MIT showing the metallic
phase\protect\index{metallic phase} and the insulating phase\protect\index{insulating phase}, respectively, at temperatures below the
critical end point, as well as a coexistence region\protect\index{coexistence region}; from Ref.~\protect\cite{Vollhardt:Bluemer-Diss}.}
\label{Vollhardt:MIT_Bluemer}
\end{figure}
As shown in Fig.~\ref{Vollhardt:MIT_Bluemer} the hysteresis region terminates  at a critical end point.
At higher temperatures
the transition changes into a smooth crossover from  a bad metal to a bad
insulator.  At half filling and for bipartite lattices in dimensions $d>2$
(in  $d=2$ only at $T=0$) the paramagnetic phase is, in fact, unstable against
antiferromagnetic long-range order. The metal-insulator transition  is then
completely hidden by the antiferromagnetic insulating phase \cite{Vollhardt:AF_Pruschke}.

In Fig.~\ref{Vollhardt:MIT_Bluemer} it is seen that the slope of the phase transition  line $U_c$
is negative down to $T=0$, which implies that for constant  interaction $U$
the metallic phase can be reached from the insulator  by decreasing the
temperature $T$, i.e., by cooling. This  anomalous behavior (which
corresponds to the Pomeranchuk effect \cite{Vollhardt:VW90}\index{Pomeranchuk effect} in $^3$He, if we
associate solid $^3$He with the  insulator and liquid $^3$He with the metal)
can be understood  from the Clausius-Clapeyron equation\index{Clausius-Clapeyron equation} $dU/dT=\Delta
S/\Delta D$. Here  $\Delta S$ is the difference between the entropy in the
metal and in  the insulator, and $\Delta D$ is the difference between the
number  of doubly occupied sites in the two phases. Within the
DMFT there is no exchange coupling $J$ between the spins of the  electrons
in the insulator, since the scaling \eqref{Vollhardt:G11.11} implies  $J \propto
-t^{2}/U \propto 1/d \rightarrow 0$ for $d\rightarrow  \infty$. Hence the
insulating state is  macroscopically degenerate, with entropy\index{entropy}  $S_{\mathrm{ins}}=k_{B}\ln2$ per electron down to  $T=0$. This is larger than the
entropy $S_{\mathrm{met}} \propto T$ per  electron in the Landau
Fermi liquid\index{Fermi liquid} describing the metal, i.e.,  $\Delta S = S_{\mathrm{met}} - S_{\mathrm{ins}} < 0$. At the same time the  number of doubly occupied sites is
lower in the insulator than in  the metal, i.e., $\Delta D = D_{\mathrm{met}} - D_{\mathrm{ins}}>0$. The  Clausius-Clapeyron equation then implies that
the phase-transition  line $T$ vs. $U$ has indeed a negative slope down to $T=0$.
However, this  is an artifact of the DMFT. Namely, there will usually  exist an exchange coupling between the electrons which
leads to a  vanishing entropy of the insulator for $T\rightarrow 0$. Since
the entropy of  the insulator vanishes faster than linearly with the
temperature,  the difference $\Delta S = S_{\mathrm{met}} - S_{\mathrm{ins}}
$  eventually becomes positive, whereby the slope also becomes positive\footnote{Here we assume for simplicity that  the metal remains a Fermi liquid and
the insulator stays  paramagnetic down to the lowest temperatures.}; this is indeed observed in cluster DMFT calculations \cite{Vollhardt:Park08}.
However, since $\Delta S\rightarrow 0$ for $T\rightarrow 0$ the phase
boundary must eventually terminate at  $T=0$ with infinite slope.

\section{Theory of Electronic Correlations in Materials}

\label{Vollhardt:Theory of Electronic Correlations in Materials}

\subsection{The LDA+DMFT Approach}

\label{Vollhardt:LDA+DMFT}

Although the Hubbard model is able to explain basic  features of the phase
diagram of  correlated electrons it cannot describe the physics of real
materials in any detail. Clearly, realistic theories must take  into account
the explicit electronic and lattice structure of the systems.

Until recently the electronic properties of solids were investigated by two
essentially separate communities, one using model Hamiltonians in
conjunction with many-body techniques, the other employing
density functional theory\index{density functional theory}
(DFT)\index{DFT|see{density functional theory}} \cite{Vollhardt:DFT1,Vollhardt:DFT2}. DFT and its
local density approximation\index{local density approximation}
(LDA)\index{LDA|see{local density approximation}}
have the advantage of being \emph{ab initio} approaches
which do not require empirical parameters as input. Indeed, they are highly
successful techniques for the calculation of the electronic structure of
real materials~\cite{Vollhardt:JonesGunn}. However, in practice DFT/LDA is seriously
restricted in its ability to describe strongly correlated materials where
the on-site Coulomb interaction is comparable with the band width. Here, the
model Hamiltonian approach is more powerful since there exist
systematic theoretical techniques to investigate the many-electron problem
with increasing accuracy. Nevertheless, the uncertainty in the choice of the
model parameters and the technical complexity of the correlation problem
itself prevent the model Hamiltonian approach from being  a suitable tool for studying real materials. The two approaches are
therefore  complementary. In view of the individual power of DFT/LDA and the
model Hamiltonian approach, respectively, a
combination of these techniques for \emph{ab initio} investigations of correlated materials including,  for example, $f$-electron systems
and Mott insulators,  would be highly desirable. One of the first successful  attempts in this direction
was the LDA+U method\index{LDA+U method} \cite{Vollhardt:Anisimov91,Vollhardt:Anisimov97}, which  combines LDA with
a static, i.e., Hartree-Fock-like, mean-field  approximation for a
multi-band Anderson lattice model  with interacting and non-interacting
orbitals. This method proved to  be a very useful tool in the study of
long-range ordered, insulating states  of transition metals and rare-earth
compounds. However, the paramagnetic  metallic phase of correlated electron
systems such as high-temperature  superconductors and heavy-fermion systems
clearly requires a treatment that  goes beyond a static mean-field
approximation and includes dynamical  effects, i.e., the frequency
dependence of the self-energy.

Here the recently developed LDA+DMFT method\index{LDA+DMFT method}, a new computational scheme
which  merges electronic band structure calculations and the DMFT, has proved to be a breakthrough \cite{Vollhardt:Anisimov97a,Vollhardt:LichtKats98,Vollhardt:Nekrasov00,Vollhardt:ldadmfta,Vollhardt:NIC2002,Vollhardt:licht,Vollhardt:psi-k,Vollhardt:dmft_phys_today,Vollhardt:Kotliar06,Vollhardt:Held07,Vollhardt:Katsnelson08,Vollhardt:Anisimov-book-2010,Vollhardt:Kunes10}. Starting  from
conventional band structure calculations in the LDA the correlations are taken into account by the  Hubbard interaction
and a Hund's rule coupling term. The resulting  DMFT equations are solved
numerically, e.g., with a quantum Monte-Carlo  (QMC) algorithm. By construction,
LDA+DMFT includes the correct  quasiparticle physics and the corresponding
energetics. It also reproduces the LDA results in the limit of weak  Coulomb
interaction $U$.  More importantly, LDA+DMFT correctly describes the
correlation induced dynamics near a Mott-Hubbard MIT and beyond.  Thus,
LDA+DMFT is able to account for the physics at all values of the  Coulomb
interaction and doping level.

In the LDA+DMFT approach  the LDA band structure is expressed  by a one-particle
Hamiltonian $\hat{H}_{\mathrm{LDA}}^{0}$, and is  then supplemented by the
local Coulomb repulsion $U$ and Hund's  rule exchange\index{Hund's  rule exchange} $J$. This leads to a material
specific  generalization of the one-band model Hamiltonian

\begin{equation}
\hat{H} = \hat{H}_{\mathrm{LDA}}^{0}+{\ U}\sum_{m}\sum_{i}\hat{n}_{im\uparrow }\hat{n}_{i m\downarrow } \;+\;\sum_{i, m\neq {m^{\prime}},
\sigma,{\sigma^{\prime}}}\;(V-\delta _{\sigma {\sigma^{\prime}}}J)\;\hat{n}_{im\sigma }\hat{n}_{im^{\prime}\sigma^{\prime}}.  \label{Vollhardt:H}
\end{equation}

Here $m$ and $m^{\prime}$ enumerate those orbitals for which the interaction between the electrons is explicitly included, e.g., the  three $t_{2g}$  orbitals
of the $3d$ electrons of transition metal ions or the $4f$ orbitals in the  case of rare earth
elements.  The interaction parameters are related by $ V=U-2J$ which holds
exactly for degenerate orbitals and is a good  approximation for $t_{2g}$ electrons. The actual values for $U$ and $V$  can be calculated by constrained LDA\index{constrained local-density approximation} \cite{Vollhardt:psi-k}.

In the one-particle part of the Hamiltonian
\begin{equation}
\hat{H}_{\mathrm{LDA}}^{0} = \hat{H}_{\mathrm{LDA}} -{\sum_{i}}\sum_{m\sigma} \Delta\epsilon_d \,\hat{n}_{im\sigma}.
\end{equation}
the energy term containing $\Delta\epsilon_d$ is a shift of the
one-particle potential of the interacting orbitals. It cancels the  Coulomb
contribution to the LDA results, and can also be calculated by  constrained
LDA \cite{Vollhardt:psi-k}.

Within the LDA+DMFT scheme the self-consistency condition  connecting the
self-energy $\Sigma $ and the Green function $G$ at  frequency $\omega$
reads: \vspace{-0.5cm}

\begin{eqnarray}
G_{qm,q^{\prime }m^{\prime }}(\omega )=\!\frac{1}{V_{B}}\int{{\ d^{3}}{k}}
\!&\left( \left[ \;\omega \bm{1}+\mu \bm{1}-H_{\mathrm{LDA}}^{0} (\bm{k} ) -\pmb{\Sigma}(\omega )\right]^{-1}\right)_{q m,q^{\prime }m^{\prime }} .&
\label{Vollhardt:Dyson}
\end{eqnarray}

Here, $\bm{1}$ is the unit matrix, $\mu$ the chemical potential, $H_{\mathrm{LDA} }^{0}(\bm{k})$ is the orbital matrix of the  LDA Hamiltonian derived,
for example, in a linearized  muffin-tin  orbital (LMTO) basis, $ \pmb{\Sigma}(\omega)$ denotes the self-energy\index{self-energy} matrix which is nonzero  only
between the interacting orbitals, and $[...]^{-1}$ implies  the inversion of
the matrix with elements $n$ (=$qm$), $n^{\prime }$(=$q^{\prime }m^{\prime }$), where $q$ and $m$ are the indices  of the atom in the primitive cell and
of the orbital,  respectively\footnote{We note that $\hat{H}_{\mathrm{LDA}}^{0}$ may
include  additional non-interacting orbitals.}. The integration extends over the Brillouin
zone with  volume $V_{B}$.

For cubic transition metal oxides  Eq.\ \eqref{Vollhardt:Dyson}  can be simplified to
\begin{eqnarray}
G(\omega)&\!=\!&G^{0}(\omega-\Sigma (\omega))=\int d\epsilon \frac{N^{0}(\epsilon )}{\omega-\Sigma (\omega)-\epsilon}  \label{Vollhardt:intg}
\end{eqnarray}
provided the degenerate $t_{2g}$ orbitals crossing the Fermi level are  well
separated from the other orbitals~\cite{Vollhardt:psi-k}. For non-cubic  systems the
degeneracy is lifted. In this case eq.\  \eqref{Vollhardt:intg} is an
approximation where different $\Sigma_m  (\omega)$, $N^{0}_m(\epsilon )$
and $G_m(\omega)$ have to be used for the three  non-degenerate  $t_{2g}$ orbitals.

The Hamiltonian \eqref{Vollhardt:H} is diagonalized within the DMFT where, for example,  quantum
Monte-Carlo (QMC) techniques~\cite{Vollhardt:Hirsch86} are used to solve the self-consistency
equations. From the imaginary time QMC Green function  we calculate the
physical (real frequency) spectral function with  the maximum entropy method~\cite{Vollhardt:MEM}.

During the last few years the LDA+DMFT and other DMFT based computational schemes have provided great progress in the understanding of the electronic, magnetic
and structural properties of many correlated electron materials. These materials
range from 3$d$ transition metals and their oxides\index{transition metal oxides}, and $f$ electron systems, all
the way to Heusler alloys, ferromagnetic half-metals, fullerenes, and zeolites \cite{Vollhardt:Kotliar06,Vollhardt:Held07,Vollhardt:Katsnelson08,Vollhardt:Anisimov-book-2010,Vollhardt:Kunes10}.
Nevertheless, this framework still needs to be considerably improved before it becomes a truly comprehensive \emph{ab initio} approach  for complex correlated matter with predictive
power. In particular, the interface between the band-structure and the
many-body components of the approach needs to be optimized. This includes,
for example, a solution of the double counting correction problem and a fully self-consistent treatment of the spin, orbital, and
charge densities. Another important goal are realistic computations of free
energies and forces, and the development of efficient methods to treat
non-local correlations with quantum cluster methods\index{quantum cluster methods} \cite{Vollhardt:Potthoff2005,Vollhardt:Maier-cluster,Vollhardt:Held08}.

\subsection{Single-Particle Spectrum of Correlated Electrons in Materials}
\label{Vollhardt:Single-Particle Spectrum of Correlated Electrons in Materials}

Transition metal oxides\index{transition metal oxides} are an ideal laboratory for the study of  electronic
correlations in solids. Among these materials, cubic  perovskites have the
simplest crystal structure and thus may be  viewed as a starting point for
understanding the electronic  properties of more complex systems. Typically,
the $3d$ states in  those materials form comparatively narrow bands with
width  $W\!\!\sim \!2\!-\!3\,$~eV, which leads to strong Coulomb
correlations between the electrons. Particularly simple are  transition
metal oxides with a 3$d^{1}$ configuration since, among  others, they do not
show a complicated multiplet structure.

Photoemission spectra provide a direct experimental tool to study  the
electronic structure and spectral properties of electronically  correlated
materials. Intensive experimental investigations of  spectral and transport
properties of strongly correlated 3$d^{1}$  transition metal oxides started
with investigations by Fujimori  \textit{et al.}~\cite{Vollhardt:fujimori}. These
authors observed a pronounced  lower Hubbard band in the photoemission
spectra  which cannot  be explained by conventional band structure
theory.
In
photoemission spectroscopy\index{photoemission spectroscopy}
(PES)\index{PES|see{photoemission spectroscopy}}
a photon of a given energy is  used to
emit an electron whose properties (energy, angular  distribution) are
measured by a detector. Angular resolved PES\index{angular resolved photoemission spectroscopy} is  referred to as ARPES.\index{ARPES|see{angular resolved photoemission spectroscopy}} These
techniques measure the \emph{occupied}  electronic states, i.e., those states which
are described by the  full spectral function multiplied by the Fermi
function $f(\omega,T)$. By contrast, inverse  photoemission spectroscopy
(IPES) measures the \emph{unoccupied}  electronic states, i.e., the states described by the full spectral function of a material multiplied by $1-f (\omega,T)$. IPES is harder to perform
and not as accurate as  PES. But in many situations information about the
unoccupied states  is also available by
X-ray absorption spectroscopy\index{X-ray absorption spectroscopy}
(XAS)\index{XAS|see{X-ray absorption spectroscopy}}.

Spectroscopic techniques provide very valuable  information about correlated
electronic systems since they can directly measure the spectral  function of
a material, a quantity which can also be directly calculated as discussed in  sec.~\ref{Vollhardt:The Mott-Hubbard
Metal-Insulator Transition}. In
particular, photoemission techniques allow one to detect the  correlation
induced shift of spectral weight .  In the following we will illustrate the
computation of the  $\bm{k}$-integrated electronic spectra of correlated
materials within the LDA+DMFT scheme by investigating the two simple transition metal oxides SrVO$_{3}$ and CaVO$_{3}$.

\subsubsection{Sr$_x$Ca$_{1-x}$VO$_3$}

SrVO$_{3}$ and CaVO$_{3}$ are simple transition metal compounds with  a 3$d^{1}$ configuration. The main effect of the substitution of Sr  ions by the
isovalent, but smaller, Ca ions is to decrease the V-O-V  angle from $\theta
= 180^{\circ }$ in SrVO$_{3}$ to $\theta \approx  162^{\circ }$ in the
orthorhombically distorted structure of  CaVO$_{3}$. Remarkably this rather
strong bond bending results only in  a 4\% decrease of the one-particle
bandwidth $W$ and thus in a  correspondingly small increase of the ratio $U/W
$ as one moves from  SrVO$_{3}$\index{SrVO3@SrVO$_{3}$} to CaVO$_{3}$\index{CaVO3@CaVO$_{3}$} \cite{Vollhardt:Sekiyama03,Vollhardt:Nekrasov05}.

\begin{figure}[tbp]
\centerline{\includegraphics[width=0.7\textwidth,clip=true]{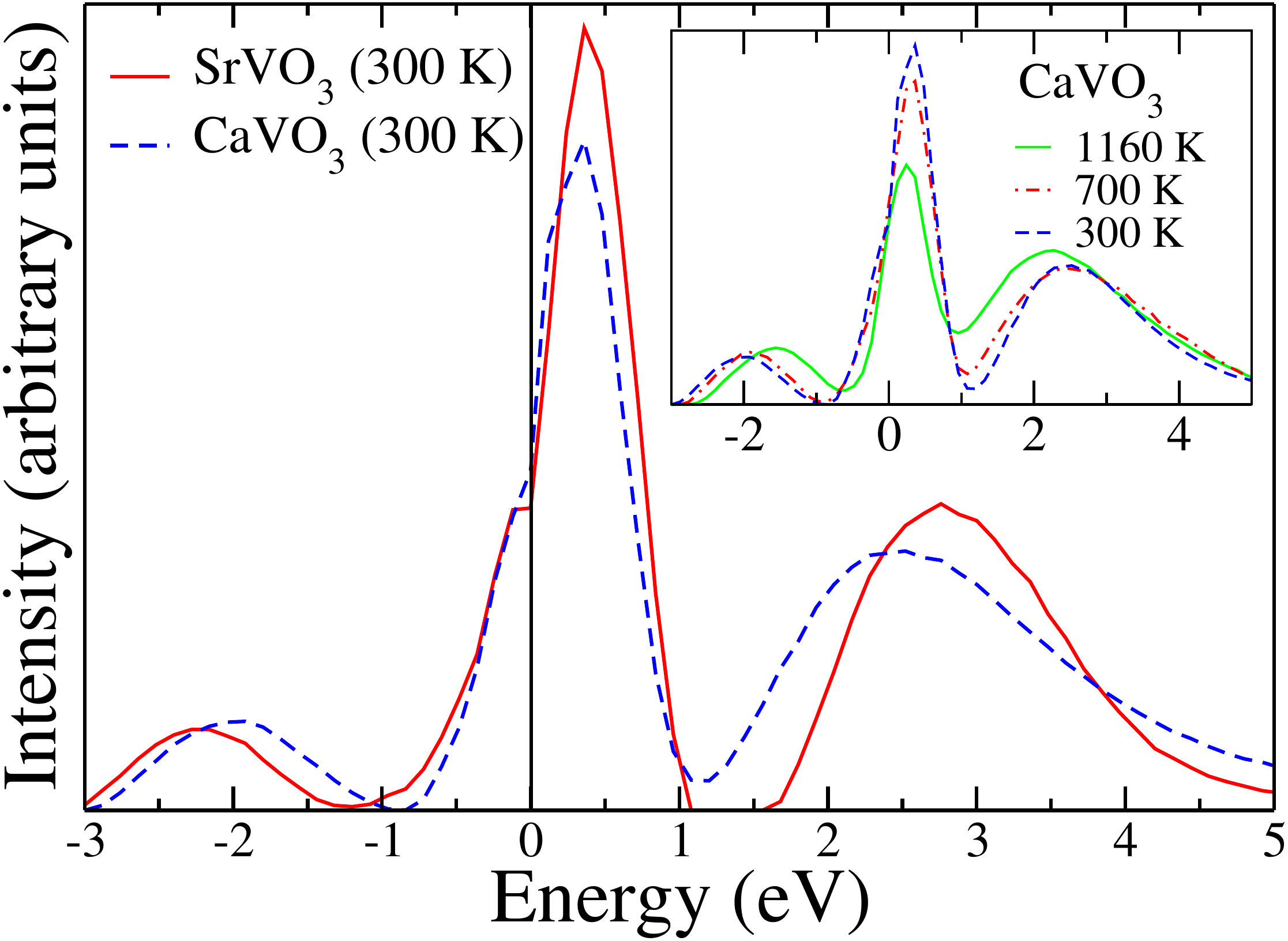}}
\caption[LDA+DMFT(QMC) spectral functions of SrVO$_{3}$ and CaVO$_{3}$.]{LDA+DMFT(QMC) spectral functions of SrVO$_{3}$ (solid line) and CaVO$_{3}$
(dashed line) calculated at T=300 K; inset: effect of temperature in the
case of CaVO$_{3}$; after Ref.~\protect\cite{Vollhardt:Sekiyama03}.}
\label{Vollhardt:fig_ldadmft}
\end{figure}

LDA+DMFT(QMC) spectral functions of SrVO$_{3}$ and CaVO$_{3}$ were calculated  by
Sekiyama \textit{et al.} \cite{Vollhardt:Sekiyama03} by starting from the  respective
LDA DOS of the two materials; they are shown in Fig.~\ref{Vollhardt:fig_ldadmft}.
These spectra show genuine correlation effects, i.e., the  formation of
lower Hubbard bands at about 1.5 eV and upper  Hubbard bands\index{Hubbard bands} at about 2.5
eV, with well-pronounced quasiparticle  peaks at the Fermi energy. Therefore
both SrVO$_{3}$ and  CaVO$_{3}$ are strongly correlated metals.
 The small
 difference of the LDA bandwidth of SrVO$_{3}$ and CaVO$_{3}$ is
 only reflected in some additional transfer of spectral weight from
 the quasiparticle peak to the Hubbard bands, and minor differences
 in the positions of the Hubbard bands.
The DOS of the two systems shown in Fig.~\ref{Vollhardt:fig_ldadmft} are  quite
similar. In fact, SrVO$_{3}$ is slightly less correlated  than CaVO$_{3}$,
in accord with their different LDA bandwidths.  The inset of Fig.~\ref{Vollhardt:fig_ldadmft} shows that the effect of  temperature on the spectrum is weak
for $T \lesssim 700$~K.
Detailed spectra of SrVO$_{3}$ and CaVO$_{3}$ were also computed by
Pavarini \emph{et al.}\cite{Vollhardt:Pavarini}.

Since the three $t_{2g}$ orbitals of this simple 3$d^{1}$ material  are
almost degenerate the spectral function has the same  three-peak
structure as that of the one-band Hubbard model shown  in Fig.~\ref{Vollhardt:fig:4.1}. The temperature induced decrease of the  quasiparticle peak height is also
clearly seen. We note that the actual form of the spectrum no longer resembles the LDA
DOS\index{density of states} used as input, i.e., it essentially depends only on the first three  energy
moments of the LDA DOS (electron density, average energy,  band width).

\begin{figure}[tbp]
\centerline{\includegraphics[width=0.9\textwidth,clip=true]{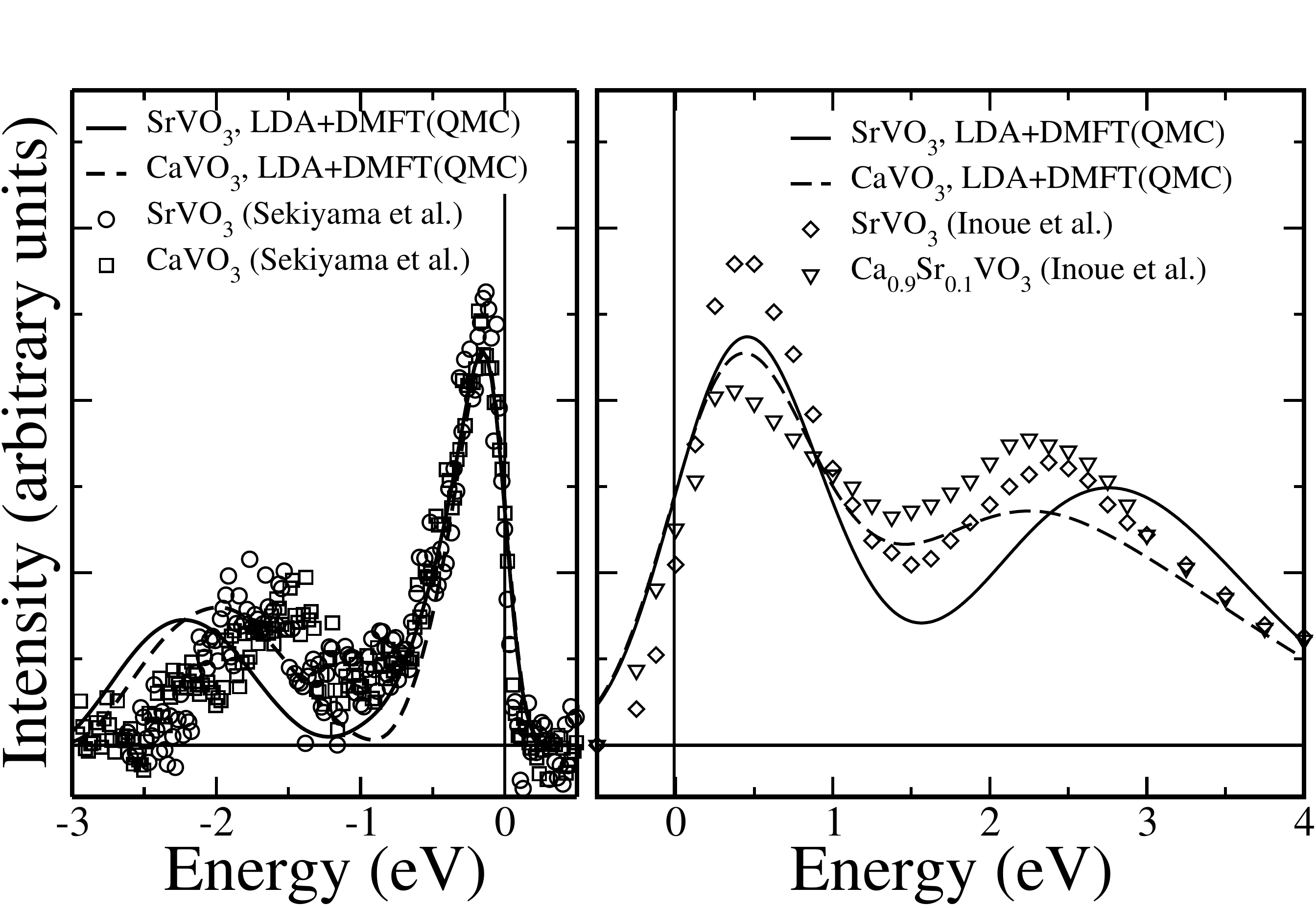}}
\caption[Comparison of LDA+DMFT(QMC) spectral functions
of SrVO$_{3}$ and CaVO$_{3}$ with experiment.]{Comparison of the calculated, parameter-free LDA+DMFT(QMC) spectral functions
for SrVO$_{3}$ (solid line) and CaVO$_{3}$ (dashed line) with experiment.
Left: Bulk-sensitive high-resolution PES (SrVO$_{3}$: circles; CaVO$_{3}$:
rectangles). Right: 1s XAS for SrVO$_{3}$ (diamonds) and Ca$_{0.9}$Sr$_{0.1}$VO$_{3}$ (triangles)~\protect\cite{Vollhardt:Inoue94}. Horizontal line: experimental
subtraction of the background intensity; after Ref.~\protect\cite{Vollhardt:Nekrasov05}.}
\label{Vollhardt:fig_XPS}
\end{figure}

In the left panel of Fig.~\ref{Vollhardt:fig_XPS} the  LDA+DMFT(QMC) spectral functions at 300K
are compared with experimental high-resolution bulk  PES.  For this purpose
the full theoretical spectra were multiplied with the  Fermi function at the
experimental temperature (20$\,$K) and were Gauss  broadened with the
experimental resolution of $0.1\,$eV  \cite{Vollhardt:Sekiyama03}. The quasiparticle
peaks in theory and experiment  are seen to be in very good agreement. In
particular, their height  and width are almost identical for both SrVO$_{3}$
and CaVO$_{3}$.  The difference in the positions of the lower Hubbard bands
may be  partly due to (i) the subtraction of the estimated oxygen
contribution (which may also remove some $3d$ spectral weight  below $-2$~eV), and (ii) uncertainties in the \emph{ab initio}  calculation of the
local Coulomb interaction strength. In the right panel of Fig.~\ref{Vollhardt:fig_XPS}
comparison is made with XAS data of Inoue \textit{et al.}~\cite{Vollhardt:Inoue94}.
To this end
the full LDA+DMFT spectrum was multiplied with the inverse Fermi function
at 80K and was then Gauss  broadened with the experimental resolution of  $0.36\,$eV~\cite{Vollhardt:Inoue03}. The overall agreement of the  weights and
positions of the quasiparticle and upper $t_{2g}$  Hubbard band is good,
including the tendencies when going from  SrVO$_{3}$ to CaVO$_{3}$ (in fact, Ca$_{0.9}
$Sr$_{0.1}$VO$_{3}$ in the  experiment). For CaVO$_{3}$ the weight of the
quasiparticle peak  is somewhat lower than in the experiment. In contrast to
one-band  Hubbard model calculations, the material specific results
reproduce the strong asymmetry around the Fermi energy w.r.t.  weights and
bandwidths.

The experimentally determined spectra of SrVO$_{3}$ and CaVO$_{3}$  and the
good agreement with parameter-free LDA+DMFT calculations  confirm the
existence of a pronounced three-peak structure in a  correlated bulk
material. Although the DMFT had predicted such a  behavior for the Hubbard
model \cite{Vollhardt:georges96} it was not clear  whether the DMFT
would really be able to describe real  materials.
Now it is clear that the  three-peak structure not only occurs in
single-impurity Anderson  models or the DMFT for the Hubbard model, but is also a characteristic feature of correlated \emph{bulk} matter in $d=3$.

\section{Electronic Correlations and Disorder}
\label{Vollhardt:Electronic Correlations and Disorder}

The properties of real materials are influenced not only by the interaction between the electrons in the periodic crystal lattice,  but also by the presence of randomness\index{randomness}, e.g., impurities and lattice defects \cite{Vollhardt:Lee85}.
 In particular, Coulomb correlations and disorder are separately driving forces behind
metal-insulator transitions\index{metal-insulator transition} (MITs) connected with the localization and
delocalization of particles.
The Mott-Hubbard MIT is caused
by the electronic repulsion \cite{Vollhardt:mott49,Vollhardt:HubbardI,Vollhardt:mott90} and is characterized by the opening of a gap in the
density of states  at the Fermi level. By contrast, the Anderson localization\index{Anderson localization}
transition is due to coherent backscattering\index{coherent backscattering}  of non-interacting particles from randomly
distributed impurities \cite{Vollhardt:Anderson58,Vollhardt:Lee85}. At the Anderson transition the character of the spectrum at the Fermi level changes from a continuous to a dense point spectrum\index{dense point spectrum}. Both MITs can be characterized by a single
quantity, the local density of states\index{local density of states} (LDOS).
 Although the LDOS is not an order parameter associated with a symmetry-breaking
phase transition, it can distinguish between a metal and an insulator.

The DMFT can easily be extended to study correlated lattice electrons with
local disorder \cite{Vollhardt:Janis_a,Vollhardt:Ulmke_aa,Vollhardt:Ulmke_ab,Vollhardt:Dobro_a,Vollhardt:Dobro_b,Vollhardt:Potthoff_a,Vollhardt:Vollhardt-Salerno}. For this purpose
a single-particle term with random local energies $\epsilon_i$ is added to the Hubbard model, leading to the Anderson-Hubbard model
\begin{equation}
\hat{H}=-t\sum_{ij,\sigma }\hat{c}_{i\sigma }^{+}
\hat{c}_{j\sigma}^{\phantom{+}}+
\sum_{i\sigma}
\epsilon_i n_{i\sigma} + U\sum_{i}\hat{n}_{i\uparrow
}\hat{n}_{i\downarrow }.  \label{Vollhardt:one}
\end{equation}
The ionic energies
$\epsilon_i$  describe the local, quenched disorder acting on the motion of the
electrons. They are drawn from a probability distribution function\index{probability distribution function} $\mathcal{P}(\epsilon_i)$,
which can be a continuous or a multi-modal function.

The DMFT provides a valuable, non-perturbative theoretical framework also for the investigation of correlated electrons in the presence of disorder.
If in the DMFT the effect of local disorder is taken into account
through the arithmetic mean\index{arithmetic mean} of the LDOS one obtains, in the
absence of interactions ($U=0$), the coherent potential approximation\index{coherent potential approximation} (CPA)
\cite{Vollhardt:vlaming92,Vollhardt:Janis92a}, which does not describe the physics of  Anderson
localization. To overcome this deficiency Dobrosavljevi\'{c}  and collaborators formulated a variant of the DMFT where the
\emph{geometrically} averaged LDOS is computed from the solutions of the
self-consistent stochastic DMFT\index{stochastic DMFT} equations
\cite{Vollhardt:Dobrosavljevic97} which is then incorporated into the self-consistency cycle \cite{Vollhardt:Dobrosavljevic03}.
Thereby a mean-field theory of Anderson localization can be derived which
reproduces many of the expected features of the disorder-driven
MIT for non-interacting electrons  \cite{Vollhardt:Dobrosavljevic03}.
This scheme  uses only one-particle quantities and can therefore
easily be included in the DMFT to treat disordered electrons in the presence of
phonons \cite{Vollhardt:fehske} or Coulomb correlations \cite{Vollhardt:Byczuk05,Vollhardt:Byczuk09_b}.
In particular, the DMFT with
geometrical averaging\index{geometrical averaging} allows one to compute the phase diagram for the
Anderson-Hubbard model \cite{Vollhardt:Byczuk05,Vollhardt:Byczuk09_b}. For a continuous disorder distribution
function 
and a  half filled band one finds
 that the metallic phase is \emph{enhanced} at small and
intermediate values of the interaction and disorder, but that the metallicity is
eventually destroyed when the disorder is strong enough \cite{Vollhardt:Byczuk05}. Surprisingly, the Mott and Anderson insulators
 are continuously connected.

New interesting phenomena are also expected in correlated electron systems when the
disorder distribution function has a binary-alloy form.
Namely, it was predicted that a disorder induced splitting of the band (``alloy-band splitting'') can enhance the critical temperature for the onset of itinerant ferromagnetism \cite{Vollhardt:Byczuk03_aa,Vollhardt:Byczuk03_ab}.
Another direct consequence
of the alloy-band splitting is the fact that, if the alloy subband is effectively
half-filled, a Mott-Hubbard metal-insulator transition\index{metal-insulator transition} can occur  at non-integer filling
\cite{Vollhardt:Byczuk04_a}. The spectral-weight transfer in correlated
systems with binary-alloy disorder\index{binary-alloy disorder} was also investigated away from the
metal-insulator transition regime \cite{Vollhardt:Lombardo_a,Vollhardt:Potthoff_07}.
For the periodic Anderson model with the binary-alloy disorder analogous behavior was predicted
\cite{Vollhardt:Unjong_a}.

\section{DMFT for Correlated Bosons in Optical Lattices}
\label{Vollhardt:DMFT for Correlated Bosons in Optical Lattices}

The observation of
Bose-Einstein condensation\index{Bose-Einstein condensation}
(BEC)\index{BEC|see{Bose-Einstein condensation}}
in ultra-cold atomic gases\index{ultra-cold atomic gases} has
greatly stimulated research into the properties of
this fascinating quantum state of matter \cite{Vollhardt:bec}. In particular,
experiments with alkali atoms confined in optical lattices\index{optical lattices}
\cite{Vollhardt:Greiner02,Vollhardt:lewenstein06,Vollhardt:Bloch07} have renewed the theoretical
interest \cite{Vollhardt:jaksch98,Vollhardt:Isacsson05,Vollhardt:Huber07} in the physics of
strongly correlated bosons on lattices, which promises significant
new insights and even applications in fields such as quantum
computing \cite{Vollhardt:micheli06}.

A lattice model of interacting bosons with purely local interaction, the  bosonic
Hubbard model,\index{bosonic Hubbard model} has the form
\begin{equation}
H_{\mathrm{BH}} = \sum_{\langle i, j\rangle } t_{ij}
\hat{b}_i^{\dagger} \hat{b}_j + \frac{U}{2} \sum _{\mathbf{R}_i} \hat{n}_i (\hat{n}_1-1),
\label{Vollhardt:B-HM}
\end{equation}
where $\hat{b}_i$ and $\hat{b}_i^{\dagger}$ are bosonic operators.
In the ground state two characteristically different phases are expected to occur: a bosonic
incompressible Mott phase with a correlation gap
 and a compressible superfluid phase characterized by a non-vanishing expectation value $\langle \hat{b}_i \rangle$ which serves as an
order parameter \cite{Vollhardt:Fisher_a}.

The construction of a DMFT for the bosonic Hubbard model
which --- as in the case of the fermionic DMFT ---
becomes exact in the limit $d$ or $Z\rightarrow \infty$ and is valid  at all temperatures  is made complicated by the fact that the system can Bose condense below a Bose-Einstein condensation temperature $T_{\mathrm{BEC}}$. This has immediate consequences for the expectation value of the kinetic energy in \eqref{Vollhardt:B-HM}. Namely, while in the normal state  only the product $\hat{b}_i^{\dagger} \hat{b}_j$ has a finite expectation value, in the Bose condensed phase even $\langle \hat{b}_j \rangle$ (the order parameter) is non-zero. For the expectation value of the kinetic energy in \eqref{Vollhardt:B-HM} to remain finite in the limit $Z\rightarrow \infty$ the hopping amplitude then has to be scaled\index{scaling of hopping amplitudes} differently in the two phases. In the normal phase the hopping amplitude needs to be scaled as in the fermionic case (``quantum scaling''), i.e., as $t_{ij}=t^*_{ij}/\sqrt{Z}$, while in the condensed phase a classical scaling $t_{ij}=t^*_{ij}/Z$ is required.
Such a scaling of the hopping amplitudes cannot be performed on the level of the Hamiltonian, but is possible in the effective action \cite{Vollhardt:Byczuk08_a}. The bosonic DMFT (``B-DMFT'')\index{bosonic DMFT} derived thereby treats normal and condensed bosons on equal footing and thus includes the effects caused by their dynamic coupling.
The self-consistency equations of the B-DMFT are those of a single bosonic impurity coupled to two baths, one corresponding to bosons in the normal state and one to bosons in the condensate.
The B-DMFT derived in the limit $d$ or $Z\rightarrow \infty$  is again a
comprehensive mean-field theory, i.e., it is valid for all input parameters and all temperatures. It not only reproduces all exactly solvable limits, such as the non-interacting ($U=0$) and the atomic ($t_{ij}=0$) limit, but also well-known approximation schemes for interacting bosons. For example, by neglecting all terms containing the hybridization function\index{hybridization function} in the local action one obtains the mean-field theory of Fisher \emph{et al.} \cite{Vollhardt:Fisher_a}; for a detailed  discussion see \cite{Vollhardt:Byczuk08_a}.

The solution of the self-consistent bosonic impurity problem defined by the B-DMFT equations requires new theoretical/computational methods. So far
exact diagonalization \cite{Vollhardt:Snoek_a,Vollhardt:Tong_a} and continuous-time QMC \cite{Vollhardt:Anders_a} were employed. Thereby the bosonic Hubbard model was solved on the Bethe lattice\index{Bethe lattice} for finite \cite{Vollhardt:Snoek_a} and infinite  \cite{Vollhardt:Tong_a}
coordination number $Z$ as well as on a simple-cubic
lattice \cite{Vollhardt:Anders_a}. The  phase diagram of correlated bosons on a simple-cubic lattice computed by the bosonic DMFT was found to agree    with that obtained by numerically exact QMC to within 2\% \cite{Vollhardt:Anders_a}.

The B-DMFT is expected  to be a valuable approximation scheme for the investigation of lattice
bosons in situations where exact numerical computations are difficult to perform or inefficient, as in the case of bosons with disorder or  many internal degrees of freedom, and for
Bose-Fermi mixtures \cite{Vollhardt:Byczuk09_a}\index{Bose-Fermi mixtures}.

\section{DMFT for Nonequilibrium}
  \label{Vollhardt:DMFT for Nonequilibrium}

  Recently the study of strongly correlated many-body systems out of
  equilibrium has received much attention
  \cite{Vollhardt:Dziarmaga2010a,Vollhardt:Polkovnikov2010a}. This is motivated in particular by
   experimental progress in the investigation of ultra-cold atomic gases\index{ultra-cold atomic gases}
  \cite{Vollhardt:Bloch07} and time-resolved pump-probe spectroscopy\index{pump-probe spectroscopy} on strongly
  correlated materials \cite{Vollhardt:Perfetti2006,Vollhardt:Wall2009}. In general the
  real-time dynamics of correlated systems can be described by the
  extension of DMFT to nonequilibrium, provided that they are
  dominated by local temporal fluctuations and spatial correlations
  are not crucial. In nonequilibrium DMFT\index{non-equilibrium DMFT} an effective impurity
  problem is formulated using the Keldysh formalism
  \cite{Vollhardt:Schmidt02,Vollhardt:Turkowski05}, and as in equilibrium DMFT this
  mapping onto a single site becomes exact in the limit of infinite
  dimensions. For the Hubbard model in nonequilibrium\index{Hubbard model!nonequilibrium@in nonequilibrium} the single-site
  DMFT action reads
  \begin{eqnarray}
    \label{Vollhardt:noneqaction}
    S
    =
    -i\int_{\mathcal{C}} \!dt \,H_{\text{loc}}(t)
    -i
    \sum_{\sigma}
    \int_{\mathcal{C}} \!dt \int_{\mathcal{C}} \!dt'\,
    c^\dagger_{\sigma}(t)
    \Lambda(t,t')
    c_{\sigma}(t')
    \,,
  \end{eqnarray}
  where the Keldysh contour $\mathcal{C}$ runs from $t_{\text{min}}$
  to $t_{\text{max}}$ on the real time axis, back to $t_{\text{min}}$,
  and finally to $-i\beta$ along the imaginary time axis
  \cite{Vollhardt:keldyshintro,Vollhardt:Eckstein2010a}.  The first term contains the
  local part of the Hamiltonian, e.g.,
  $H_{\text{loc}}(t)=U(t)n_{\uparrow}n_{\downarrow} -\mu
  (n_{\uparrow}+n_{\downarrow})$ for a time-dependent interaction. The
  second term involves the hybridization function\index{hybridization function} $\Lambda(t,t')$
  which couples the impurity to a time-dependent bath which must be
  determined self-consistently. Local con\-tour-or\-dered correlation
  functions such as the Green function\index{Green function} $G(t,t')$ are obtained from the
  action (\ref{Vollhardt:noneqaction}) as expectation values $\langle
  A(t)B(t')\cdots\rangle$ $=$
  $\text{Tr}[\text{T}_{\mathcal{C}}\exp(S)A(t)B(t')\cdots]/{\mathcal{Z}}$
  at the appropriate times, where $\text{T}_{\mathcal{C}}$ is the
  con\-tour-or\-dering operator.  For the Hubbard model this
  evaluation is the most demanding part of nonequilibrium DMFT and can
  so far be done with real-time quantum Monte Carlo
  methods 
  \cite{Vollhardt:Eckstein2009,Vollhardt:Eckstein2010a} for not too long times, and for
  sufficiently large $U$ using a self-consistent perturbation
  expansion around the atomic limit \cite{Vollhardt:Eckstein2010e}.  For the
  Falicov-Kimball model,\index{Falicov-Kimball model} on the other hand, closed equations of motion
  govern the impurity Green function \cite{Vollhardt:Brandt}, which can be solved
  on the real time axis.  The hybridization function $\Lambda(t,t')$
  in eq.~\eqref{Vollhardt:noneqaction} must be determined self-consistently by
  computing the local self-energy\index{self-energy} $\Sigma(t,t')$ from the Dyson
  equation\index{Dyson equation} of the impurity model, calculating the momentum-dependent
  Green function of the lattice model from the lattice Dyson equation,
  integrating over momentum to obtain the local lattice Green
  function, and finally equating it with the impurity Green function.
  While this procedure is necessary for a Gaussian or other general
  DOS, for a semielliptical DOS the self-energy can be eliminated and
  $\Lambda$ expressed directly in terms of $G$
  \cite{Vollhardt:Eckstein2010epjst}.

  Nonequilibrium DMFT was used to obtain the response of time-resolved
  photoemission \cite{Vollhardt:Freericks09,Vollhardt:Eckstein2008c,Vollhardt:Moritz2010a} and
  optical spectroscopy \cite{Vollhardt:Eckstein2008b} in correlated systems in
  terms of Green functions of the electronic system. The
  Falicov-Kimball and Hubbard models were studied in the presence of
  dc and ac electric fields
  \cite{Vollhardt:Turkowski05,Vollhardt:Freericks06b,Vollhardt:Tran08x,Vollhardt:Freericks08,Vollhardt:Joura2008a,Vollhardt:Tsuji08,Vollhardt:Tsuji2009a,Vollhardt:Eckstein2010d,Vollhardt:Tsuji2010a,Vollhardt:Moritz2010a},
  as well as for abrupt
  \cite{Vollhardt:Eckstein2008a,Vollhardt:Eckstein2008b,Vollhardt:Eckstein2008c,Vollhardt:Eckstein2009,Vollhardt:Eckstein2010a,Vollhardt:Eckstein2010e}
  or slow changes \cite{Vollhardt:EcksteinNJP2010,Vollhardt:Eurich2010a} of the
  interaction parameter as a function of time.

  As an example, we consider a sudden change (``quench'')\index{quench} in the
  interation parameter of the Hubbard model from $U$ $=$ 0 (i.e., with
  the non-interacting ground state as initial state) to finite values
  of $U$ for times $t$ $>$ $0$.  For this case the DMFT equations were
  solved numerically for the paramagnetic phase and a semielliptic DOS
  \cite{Vollhardt:Eckstein2009,Vollhardt:Eckstein2010a,Vollhardt:Eckstein2010e}. In the following
  discussion the bandwidth is equal to $4$, and time is thus measured
  in units of $4\hbar/\text{bandwidth}$, e.g., on the order of
  femto-seconds for a bandwidth on the order of eV.  An interesting
  question is whether such an isolated system can thermalize\index{thermalization} due to
  the many-body interaction alone, i.e., whether at least some of its
  properties are the same as for an equilibrium system with the same
  energy~\cite{Vollhardt:Rigol2008a}. Before the quench the momentum
  distribution is a step function, and the Fermi surface discontinuity\index{Fermi surface discontinuity}
  $\Delta n(t)$ remains nonzero for a finite time after the quench.
  For quenches to $U$ $\lesssim$ $2.5$ the momentum discontinuity
  first reaches a so-called prethermalization\index{prethermalization} plateau for $t$
  $\lesssim$ 5 due to the vicinity of the integrable point at $U=0$.
  This plateau in $\Delta n(t)$ is given to good accuracy by
  $2Z-1$~\cite{Vollhardt:Eckstein2009}, where $Z$ is the Fermi-liquid\index{Fermi liquid} quasiparticle weight
  \emph{in equilibrium} at zero temperature and for interaction $U$.
  This value and also the transient behavior at short times is
  precisely predicted by second-order unitary perturbation theory in
  $U$~\cite{Vollhardt:moeckel08,Vollhardt:moeckel09}.  On the other hand, the double
  occupation essentially relaxes to its thermal value on this
  timescale, showing that the potential energy (and therefore also the
  kinetic energy) relax quicker than the occupation of individual
  states.  For large $U$ the behavior is different, showing strong
  so-called collapse-and-revival oscillations\index{collapse-and-revival oscillations} with approximate
  frequency $2\pi/U$. They stem from the vicinity of the atomic limit
  (i.e., zero hopping amplitude), for which the propagator $e^{-iHt}$
  is exactly periodic with period $2\pi/U$~\cite{Vollhardt:Greiner}. For finite
  hopping (small compared to $U$) these oscillations are damped and
  decay on timescales of order $\hbar/\text{bandwidth}$.  The
  oscillations of the double occupation are not centered at its
  thermal value, but rather at a different value that can be derived
  from strong-coupling perturbation theory \cite{Vollhardt:Eckstein2009}. The
  situation is thus similar to that at small coupling in the sense
  that the relaxation to the thermal state is delayed because the
  system is stuck in a metastable state close to an integrable point.
  However, both the weak-coupling prethermalization plateau in $\Delta
  n(t)$ as well as the strong-coupling oscillations vanish in a narrow
  region of interaction parameters $U$ near $3.2$ \cite{Vollhardt:Eckstein2009}.
  For quenches of $U$ to approximately this value the system
  thermalizes rapidly: Both the momentum distribution and thus also
  the double occupation (due to energy conservation after quench)
  relax to their thermal values. In fact the retarded nonequilibrium
  Green function\index{Green function!thermalization of the} relaxes to the corresponding equilibrium
  function~\cite{Vollhardt:Eckstein2010a}, so that all observables that can be
  obtained from it tend to the thermal value predicted by equilibrium
  statistical mechanics. In other words, for quenches to interaction
  values in the vicinity of $U$ $\approx$ $3.2$ the isolated system
  indeed thermalizes rapidly due to the many-body interactions. Many
  interesting questions remain open in this context, e.g., how
  thermalization depends on the parameters of the system. We refer to
  the reviews~\cite{Vollhardt:Dziarmaga2010a,Vollhardt:Polkovnikov2010a} for further
  discussion.

\section{Summary and Outlook}
\label{Vollhardt:Summary and Outlook}

Due to the intensive international research over the last two  decades the
DMFT has quickly developed into a powerful method for  the investigation of
electronic systems with strong correlations. It  provides a comprehensive,
non-perturbative and thermodynamically  consistent approximation scheme for
the investigation of  finite-di\-men\-sion\-al systems (in particular for
dimension $d=3$), and  is particularly useful for the study of problems
where perturbative  approaches are inapplicable. For this reason the DMFT
has now become  the standard mean-field theory for fermionic correlation
problems,  including cold atoms in optical lattices~\cite{Vollhardt:Rapp,Vollhardt:Snoek,Vollhardt:Mott2}.
The study of models in nonequilibrium using an appropriate
generalization of DMFT has become yet another fascinating new research
area~\cite{Vollhardt:Turkowski05,Vollhardt:Eckstein2010a,Vollhardt:Eckstein2009,Vollhardt:Eckstein2010e,Vollhardt:Eckstein2008c,Vollhardt:Moritz2010a,Vollhardt:Eckstein2008b,Vollhardt:Freericks06b,Vollhardt:Tran08x,Vollhardt:Freericks08,Vollhardt:Joura2008a,Vollhardt:Tsuji08,Vollhardt:Tsuji2009a,Vollhardt:Eckstein2010d,Vollhardt:Tsuji2010a,Vollhardt:Eckstein2008a,Vollhardt:EcksteinNJP2010,Vollhardt:Eurich2010a}.

Until a few years ago research into correlated electron systems
concentrated on homogeneous bulk systems. DMFT investigations of  systems
with internal or external inhomogeneities such as thin films  and
multi-layered nanostructures are still very new  ~\cite{Vollhardt:Freericks-book,Vollhardt:Potthoff99,Vollhardt:Takizawa06,Vollhardt:Chen07,Vollhardt:Byczuk08x,Vollhardt:Helmes08,Vollhardt:Snoek}.
They are particularly important in view of the novel types of
functionalities of such systems, which may have important  applications in
electronic devices. Here the DMFT and its non-local extensions \cite{Vollhardt:Potthoff2005,Vollhardt:Maier-cluster,Vollhardt:Held08} will certainly become very useful.

In particular, the development of the \emph{ab initio}  band-structure
calculation technique referred to as LDA+DMFT has  proved to be a
breakthrough in the investigation of electronically  correlated materials.
It  provided already important insights  into the spectral and magnetic
properties of correlated electron  materials
\cite{Vollhardt:licht,Vollhardt:psi-k,Vollhardt:Kotliar06,Vollhardt:Held07,Vollhardt:Katsnelson08,Vollhardt:Anisimov-book-2010}. Clearly, this  approach has
a great potential for further developments. Indeed, it  is not hard to
foresee that the LDA+DMFT framework will eventually  develop into a
comprehensive \emph{ab initio} approach which is able  to describe, and even
predict, the properties of complex correlated  materials.

\section*{Acknowledgments}

We thank Jim Allen, Vladimir Anisimov, Nils Bl\"{u}mer, Ralf Bulla, Liviu Chioncel, Theo Costi, Vlad Dobrosavljevi\'{c}, Peter van Dongen, Martin Eckstein, Volker Eyert, Florian Gebhard, Karsten Held, Walter Hofstetter, Vaclav Jani\v{s}, Anna Kauch, Stefan Kehrein, Georg Keller, Gabi Kotliar, Jan Kune\v{s}, Ivan Leonov, Walter Metzner, Michael Moeckel, Igor Nekrasov, Thomas Pruschke, Xinguo Ren, Shigemasa Suga, G\"{o}tz Uhrig, Martin Ulmke, Ruud Vlaming, Philipp Werner, and Unjong Yu for valuable collaborations. Support by the Deutsche Forschungsgemeinschaft through  TRR~80 and FOR~1346 is gratefully acknowledged. KB was also supported by the grant N~N202~103138 of the Polish Ministry of Science and Education.

\hyphenation{Post-Script Sprin-ger}


\begin{thebibliography}{100}

\bibitem{Vollhardt:deboer}
J.H. de~Boer, E.J.W. Verwey, Proc. Phys. Soc. \textbf{49}, 59 (1937)

\bibitem{Vollhardt:mott37}
N.F. Mott, R.~Peierls, Proc. Phys. Soc. A \textbf{49}, 72 (1937)

\bibitem{Vollhardt:tokura}
M.~Imada, A.~Fujimori, Y.~Tokura, Rev. Mod. Phys. \textbf{70}, 1039 (1998)

\bibitem{Vollhardt:Gutzwiller}
M.C. Gutzwiller, Phys. Rev. Lett. \textbf{10}, 59 (1963)

\bibitem{Vollhardt:HubbardI}
J.~Hubbard, Proc. Roy. Soc. London \textbf{A276}, 238 (1963)

\bibitem{Vollhardt:Kanamori}
J.~Kanamori, Prog. Theor. Phys. \textbf{30}, 275 (1963)

\bibitem{Vollhardt:Lieb+Wu}
E.~Lieb, F.Y. Wu, Phys. Rev. Lett. \textbf{20}, 1445 (1968)

\bibitem{Vollhardt:Vollhardt-Salerno}
D.~Vollhardt, in \emph{Lectures on the Physics of Strongly Correlated Systems
  {XIV}}, \emph{AIP Conference Proceedings}, vol. 1297, ed. by A.~Avella,
  F.~Mancini (American Institute of Physics, Melville, 2010), p. 339

\bibitem{Vollhardt:Jerusalem}
D.~Vollhardt, in \emph{Correlated Electron Systems}, ed. by V.J. Emery (World
  Scientific, Singapore, 1993), p.~57

\bibitem{Vollhardt:Baxter}
R.J. Baxter, \emph{Exactly Solved Models in Statistical Mechanics} (Academic
  Press, London, 1982)

\bibitem{Vollhardt:Itzykson89}
C.~Itzykson, J.M. Drouffe, \emph{Statistical Field Theory} (Cambridge
  University Press, Cambridge, 1989)

\bibitem{Vollhardt:metzner89}
W.~Metzner, D.~Vollhardt, Phys. Rev. Lett. \textbf{62}, 324 (1989)

\bibitem{Vollhardt:Wolff83}
U.~Wolff, Nucl. Phys. B \textbf{225}, 391 (1983)

\bibitem{Vollhardt:MH89a}
E.~M{\"u}ller-Hartmann, Z. Phys. B \textbf{74}, 507 (1989)

\bibitem{Vollhardt:metzner89a}
W.~Metzner, Z. Phys. B \textbf{77}, 253 (1989)

\bibitem{Vollhardt:MH89b}
E.~M{\"u}ller-Hartmann, Z. Phys. B \textbf{76}, 211 (1989)

\bibitem{Vollhardt:Czycholl__1of4}
H.~Schweitzer, G.~Czycholl, Solid State Comm. \textbf{69}, 171 (1989)

\bibitem{Vollhardt:Czycholl__4of4}
H.~Schweitzer, G.~Czycholl, Z. Phys. B \textbf{83}, 93 (1991)

\bibitem{Vollhardt:Brandt}
U.~Brandt, C.~Mielsch, Z. Phys. B \textbf{75}, 365 (1989)

\bibitem{Vollhardt:vanDongen89}
P.G.J. van Dongen, F.~Gebhard, D.~Vollhardt, Z. Phys. \textbf{76}, 199 (1989)

\bibitem{Vollhardt:Kajzar78}
F.~Kajzar, J.~Friedel, J. Phys. (Paris) \textbf{39}, 397 (1978)

\bibitem{Vollhardt:Treglia80}
G.~Treglia, F.~Ducastelle, D.~Spanjaard, Phys. Rev. B \textbf{22}, 6472 (1980)

\bibitem{Vollhardt:Bulk90}
G.~Bulk, R.J. Jelitto, Phys. Rev. B \textbf{41}, 413 (1990)

\bibitem{Vollhardt:LW60}
J.M. Luttinger, J.C. Ward, Phys. Rev. \textbf{118}, 1417 (1960)

\bibitem{Vollhardt:Janis91}
V.~Jani\v{s}, Z. Phys. B \textbf{83}, 227 (1991)

\bibitem{Vollhardt:Janis92b}
V.~Jani\v{s}, D.~Vollhardt, Int. J. Mod. Phys. B \textbf{6}, 731 (1992)

\bibitem{Vollhardt:Velicky68}
B.~Velick\'y, S.~Kirkpatrick, H.~Ehrenreich, Phys. Rev. \textbf{175}, 745
  (1968)

\bibitem{Vollhardt:Yonezawa73}
F.~Yonezawa, K.~Morigaki, Suppl. Prog. Theor. Phys. \textbf{53}, 1 (1973)

\bibitem{Vollhardt:Elliot74}
R.J. Elliot, J.A. Krumhansl, P.L. Leath, Rev. Mod. Phys. \textbf{46}, 465
  (1974)

\bibitem{Vollhardt:vlaming92}
R.~Vlaming, D.~Vollhardt, Phys. Rev. B \textbf{45}, 4637 (1992)

\bibitem{Vollhardt:Georges92}
A.~Georges, G.~Kotliar, Phys. Rev. B \textbf{45}, 6479 (1992)

\bibitem{Vollhardt:Jarrell92}
M.~Jarrell, Phys. Rev. Lett. \textbf{69}, 168 (1992)

\bibitem{Vollhardt:georges96}
A.~Georges, G.~Kotliar, W.~Krauth, M.J. Rozenberg, Rev. Mod. Phys. \textbf{68},
  13 (1996)

\bibitem{Vollhardt:Fabrizio}
M.~Fabrizio, in \emph{Lectures on the Physics of highly correlated electron
  systems {XI}}, \emph{AIP Conference Proceedings}, vol. 918, ed. by A.~Avella,
  F.~Mancini (American Institute of Physics, Melville, 2007), p.~3

\bibitem{Vollhardt:Hirsch86}
J.E. Hirsch, R.M. Fye, Phys. Rev. Lett. \textbf{56}, 2521 (1986)

\bibitem{Vollhardt:fk_dmft}
J.K. Freericks, V.~Zlati{\'c}, Rev. Mod. Phys. \textbf{75}, 1333 (2003)

\bibitem{Vollhardt:Bulla}
R.~Bulla, Phys. Rev. Lett. \textbf{83}, 136 (1999)

\bibitem{Vollhardt:NRG-RMP}
R.~Bulla, T.A. Costi, T.~Pruschke, Rev. Mod. Phys. \textbf{80}, 395 (2008)

\bibitem{Vollhardt:dmrg}
M.~Karski, C.~Raas, G.S. Uhrig, Phys. Rev. B \textbf{77}, 075116 (2008)

\bibitem{Vollhardt:Caffarel94}
M.~Caffarel, W.~Krauth, Phys. Rev. Lett. \textbf{72}, 1545 (1994)

\bibitem{Vollhardt:Si94}
Q.~Si, M.J. Rozenberg, G.~Kotliar, A.E. Ruckenstein, Phys. Rev. Lett.
  \textbf{72}, 2761 (1994)

\bibitem{Vollhardt:Rozenberg94a}
M.J. Rozenberg, G.~Moeller, G.~Kotliar, Mod. Phys. Lett. B \textbf{8}, 535
  (1994)

\bibitem{Vollhardt:Rozenberg92}
M.J. Rozenberg, X.Y. Zhang, G.~Kotliar, Phys. Rev. Lett. \textbf{69}, 1236
  (1992)

\bibitem{Vollhardt:Georges92b}
A.~Georges, W.~Krauth, Phys. Rev. Lett. \textbf{69}, 1240 (1992)

\bibitem{Vollhardt:Rubtsov}
A.N. Rubtsov, V.V. Savkin, A.I. Lichtenstein, Phys. Rev. B \textbf{72}, 035122
  (2005)

\bibitem{Vollhardt:Werner}
P.~Werner, A.~Comanac, L.~de' Medici, M.~Troyer, A.J. Millis, Phys. Rev. Lett.
  \textbf{97}, 076405 (2006)

\bibitem{Vollhardt:Haule}
K.~Haule, Phys. Rev. B \textbf{75}, 155113 (2007)

\bibitem{Vollhardt:IPT}
X.Y. Zhang, M.J. Rozenberg, G.~Kotliar, Phys. Rev. Lett. \textbf{70}, 1666
  (1993)

\bibitem{Vollhardt:psi-k}
K.~Held, I.A. Nekrasov, G.~Keller, V.~Eyert, N.~Bl{\"u}mer, A.K. McMahan, R.T.
  Scalettar, T.~Pruschke, V.I. Anisimov, D.~Vollhardt, Psi-k Newsletter
  \textbf{56}, 65 (2003).
\newblock Reprinted in {Phys. Status Solidi B} \textbf{243}, 2599 (2006)

\bibitem{Vollhardt:Bickers89}
N.E. Bickers, D.J. Scalapino, Ann. Phys. (NY) \textbf{193}, 206 (1989)

\bibitem{Vollhardt:LichtKats99}
A.I. Lichtenstein, M.I. Katsnelson, J. Phys. Condens. Matter \textbf{11}, 1037
  (1999)

\bibitem{Vollhardt:Kotliar06}
G.~Kotliar, S.Y. Savrasov, K.~Haule, V.S. Oudovenko, O.~Parcollet, C.A.
  Marianetti, Rev. Mod. Phys. \textbf{78}, 865 (2006)

\bibitem{Vollhardt:Drchal05}
V.~Drchal, V.~Jani\v{s}, J.~Kudrnovsk\'y, V.S. Oudovenko, X.~Dai, K.~Haule,
  G.~Kotliar, J. Phys.: Condens. Matter \textbf{17}, 61 (2005)

\bibitem{Vollhardt:Logan}
D.E. Logan, M.T. Glossop, J. Phys. Condens. Matter \textbf{12}, 985 (2000)

\bibitem{Vollhardt:Kauch}
A.~Kauch, K.~Byczuk, ArXiv:0912.4278  (2009)

\bibitem{Vollhardt:parquet}
V.~Jani\ifmmode~\check{s}\else \v{s}\fi{}, P.~Augustinsk\'y, Phys. Rev. B
  \textbf{75}, 165108 (2007)

\bibitem{Vollhardt:freericks95}
T.~Pruschke, M.~Jarrell, J.K. Freericks, Adv. Phys. \textbf{44}, 187 (1995)

\bibitem{Vollhardt:Georges2003}
A.~Georges, in \emph{Lectures on the Physics of highly correlated electron
  systems {VIII}}, \emph{AIP Conference Proceedings}, vol. 715, ed. by
  A.~Avella, F.~Mancini (American Institute of Physics, Melville, 2004), p.~3

\bibitem{Vollhardt:dmft_phys_today}
G.~Kotliar, D.~Vollhardt, Physics Today \textbf{57}(No. 3 (March)), 53 (2004)

\bibitem{Vollhardt:mott68}
N.F. Mott, Rev. Mod. Phys. \textbf{40}, 677 (1968)

\bibitem{Vollhardt:mott90}
N.F. Mott, \emph{Metal-Insulator Transitions}, 2nd edn. (Taylor and Francis,
  London, 1990)

\bibitem{Vollhardt:Gebhard}
F.~Gebhard, \emph{The {M}ott Metal-Insulator Transition} (Springer, Berlin,
  1997)

\bibitem{Vollhardt:McWhR}
D.B. McWhan, J.P. Remeika, Phys. Rev. B \textbf{2}, 3734 (1970)

\bibitem{Vollhardt:McWhetal}
D.B. McWhan, A.~Menth, J.P. Remeika, W.F. Brinkman, T.M. Rice, Phys. Rev. B
  \textbf{7}, 1920 (1973)

\bibitem{Vollhardt:RMcWh}
T.M. Rice, D.B. McWhan, IBM J. Res. Develop. \textbf{14}, 251 (1970)

\bibitem{Vollhardt:Hub3}
J.~Hubbard, Proc. Roy. Soc. London \textbf{A281}, 401 (1964)

\bibitem{Vollhardt:BR}
W.F. Brinkman, T.M. Rice, Phys. Rev. B \textbf{2}, 4302 (1970)

\bibitem{Vollhardt:byczuk07}
K.~Byczuk, M.~Kollar, K.~Held, Y.F. Yang, I.A. Nekrasov, T.~Pruschke,
  D.~Vollhardt, Nature Physics \textbf{3}, 168 (2007)

\bibitem{Vollhardt:Toschi09}
A.~Toschi, M.~Capone, C.~Castellani, K.~Held, Phys. Rev. Lett. \textbf{102},
  076402 (2009)

\bibitem{Vollhardt:Uhrig09}
C.~Raas, P.~Grete, G.S. Uhrig, Phys. Rev. Lett. \textbf{102}, 076406 (2009)

\bibitem{Vollhardt:Lanzara2002}
A.~Lanzara, P.V. Bogdanov, X.J. Zhou, S.A. Kellar, D.L. Feng, E.D. Lu,
  T.~Yoshida, H.~Eisaki, A.~Fujimori, K.~Kishio, J.I. Shimoyama, T.~Noda,
  S.~Uchida, Z.~Hussain, Z.X. Shen, Nature \textbf{412}, 510 (2001)

\bibitem{Vollhardt:Shen2002}
Z.X. Shen, A.~Lanzara, S.~Ishihara, N.~Nagaosa, Philos. Mag. B \textbf{82},
  1349 (2002)

\bibitem{Vollhardt:He2001}
H.~He, Y.~Sidis, P.~Bourges, G.D. Gu, A.~Ivanov, N.~Koshizuka, B.~Liang, C.T.
  Lin, L.P. Regnault, E.~Schoenherr, B.~Keimer, Phys. Rev. Lett. \textbf{86},
  1610 (2001)

\bibitem{Vollhardt:Hwang2004}
J.~Hwang, T.~Timusk, G.D. Gu, Nature \textbf{427}, 714 (2004)

\bibitem{Vollhardt:Nickel09}
A.~Hofmann, X.Y. Cui, J.~Sch{\"a}fer, S.~Meyer, P.~H{\"o}pfner, C.~Blumenstein,
  M.~Paul, L.~Patthey, E.~Rotenberg, J.~B{\"u}nemann, F.~Gebhard, T.~Ohm,
  W.~Weber, R.~Claessen, Phys. Rev. Lett. \textbf{102}, 187204 (2009)

\bibitem{Vollhardt:Bulla01}
R.~Bulla, T.A. Costi, D.~Vollhardt, Phys. Rev. B \textbf{64}, 045103 (2001)

\bibitem{Vollhardt:Mo04}
S.K. Mo, H.D. Kim, J.W. Allen, G.H. Gweon, J.D. Denlinger, J.H. Park,
  A.~Sekiyama, A.~Yamasaki, S.~Suga, P.~Metcalf, K.~Held, Phys. Rev. Lett.
  \textbf{93}, 076404 (2004)

\bibitem{Vollhardt:Roz99}
M.J. Rozenberg, R.~Chitra, G.~Kotliar, Phys. Rev. Lett. \textbf{83}, 3498
  (1999)

\bibitem{Vollhardt:Joo}
J.~Joo, V.~Oudovenko, Phys. Rev. B \textbf{64}, 193102 (2001)

\bibitem{Vollhardt:Bluemer-Diss}
N.~Bl{\"u}mer, \emph{Metal-Insulator Transition and Optical Conductivity in
  High Dimensions} (Shaker Verlag, Aachen, 2003)

\bibitem{Vollhardt:AF_Pruschke}
T.~Pruschke, Prog. Theor. Phys. Suppl. \textbf{160}, 274 (2005)

\bibitem{Vollhardt:VW90}
D.~Vollhardt, P.~W{\"o}lfle, \emph{The Superfluid Phases of {H}elium 3} (Taylor
  and Francis, London, 1990)

\bibitem{Vollhardt:Park08}
H.~Park, K.~Haule, G.~Kotliar, Phys. Rev. Lett. \textbf{101}, 186403 (2008)

\bibitem{Vollhardt:DFT1}
P.~Hohenberg, W.~Kohn, Phys. Rev. \textbf{136B}, 864 (1964)

\bibitem{Vollhardt:DFT2}
W.~Kohn, L.J. Sham, Phys. Rev. \textbf{140}, A1133 (1965)

\bibitem{Vollhardt:JonesGunn}
R.O. Jones, O.~Gunnarsson, Rev. Mod. Phys. \textbf{61}, 689 (1989)

\bibitem{Vollhardt:Anisimov91}
V.I. Anisimov, J.~Zaanen, O.K. Andersen, Phys. Rev. B \textbf{44}, 943 (1991)

\bibitem{Vollhardt:Anisimov97}
V.I. Anisimov, F.~Aryasetiawan, A.I. Lichtenstein, J. Phys.: Cond. Matter
  \textbf{9}, 767 (1997)

\bibitem{Vollhardt:Anisimov97a}
V.I. Anisimov, A.I. Poteryaev, M.A. Korotin, A.O. Anokhin, G.~Kotliar, J.
  Phys.: Cond. Matt. \textbf{9}, 7359 (1997)

\bibitem{Vollhardt:LichtKats98}
A.I. Lichtenstein, M.I. Katsnelson, Phys. Rev. B \textbf{57}, 6884 (1998)

\bibitem{Vollhardt:Nekrasov00}
I.A. Nekrasov, K.~Held, N.~Bl{\"u}mer, A.I. Poteryaev, V.I. Anisimov,
  D.~Vollhardt, Eur. Phys. J. B \textbf{18}, 55 (2000)

\bibitem{Vollhardt:ldadmfta}
K.~Held, I.A. Nekrasov, N.~Bl{\"u}mer, V.I. Anisimov, D.~Vollhardt, Int. J.
  Mod. Phys. B \textbf{15}, 2611 (2001)

\bibitem{Vollhardt:NIC2002}
K.~Held, I.A. Nekrasov, G.~Keller, V.~Eyert, N.~Bl{\"u}mer, A.K. McMahan,
  R.~Scalettar, T.~Pruschke, V.~Anisimov, D.~Vollhardt, in \emph{Quantum
  Simulations of Complex Many-Body Systems: From Theory to Algorithms},
  \emph{NIC Series Volume}, vol.~10, ed. by J.~Grotendorst, D.~Marks,
  A.~Muramatsu (NIC Directors, Forschungszentrum J{\"u}lich, Berlin, 2002), p.
  175

\bibitem{Vollhardt:licht}
A.I. Lichtenstein, M.I. Katsnelson, G.~Kotliar, in \emph{Electron Correlations
  and Materials Properties}, ed. by A.~Gonis, N.~Kioussis, M.~Ciftan (Kluwer
  Academic/Plenum, New York, 2002), p. 428

\bibitem{Vollhardt:Held07}
K.~Held, Adv. Phys. \textbf{56}, 829 (2007)

\bibitem{Vollhardt:Katsnelson08}
M.I. Katsnelson, V.Y. Irkhin, L.~Chioncel, A.I. Lichtenstein, R.A. de~Groot,
Rev. Mod. Phys. \textbf{80}, 315 (2008)

\bibitem{Vollhardt:Anisimov-book-2010}
V.  Anisimov and Y. Izyumov, \emph{Electronic Structure of Correlated Materials}, Springer Series in Solid-State Sciences \textbf{163} (Springer, Berlin, 2010)


\bibitem{Vollhardt:Kunes10}
J.~Kune\v{s}, I.~Leonov, M.~Kollar, K.~Byczuk, V.I. Anisimov, D.~Vollhardt,
  Eur. Phys. J. Special Topics \textbf{180}, 5 (2010)

\bibitem{Vollhardt:MEM}
M.~Jarrell, J.E. Gubernatis, Phys. Rep. \textbf{269}, 133 (1996)

\bibitem{Vollhardt:Potthoff2005}
M.~Potthoff, Adv. Solid State Phys. \textbf{45}, 135 (2005)

\bibitem{Vollhardt:Maier-cluster}
T.~Maier, M.~Jarrell, T.~Pruschke, M.H. Hettler, Rev. Mod. Phys. \textbf{77},
  1027 (2005)

\bibitem{Vollhardt:Held08}
K.~Held, A.A. Katanin, A.~Toschi, Prog. Theor. Phys. Suppl. \textbf{176}, 117
  (2008)

\bibitem{Vollhardt:fujimori}
A.~Fujimori, I.~Hase, H.~Namatame, Y.~Fujishima, Y.~Tokura, H.~Eisaki,
  S.~Uchida, K.~Takegahara, F.M.F. de~Groot, Phys. Rev. Lett. \textbf{69}, 1796
  (1992)

\bibitem{Vollhardt:Sekiyama03}
A.~Sekiyama, H.~Fujiwara, S.~Imada, S.~Suga, H.~Eisaki, S.I. Uchida,
  K.~Takegahara, H.~Harima, Y.~Saitoh, I.A. Nekrasov, G.~Keller, D.E. Kondakov,
  A.V. Kozhevnikov, T.~Pruschke, K.~Held, D.~Vollhardt, V.I. Anisimov, Phys.
  Rev. Lett. \textbf{93}, 156402 (2004)

\bibitem{Vollhardt:Nekrasov05}
I.A. Nekrasov, G.~Keller, D.E. Kondakov, A.V. Kozhevnikov, T.~Pruschke,
  K.~Held, D.~Vollhardt, V.I. Anisimov, Phys. Rev. B \textbf{72}, 155106 (2005)

\bibitem{Vollhardt:Pavarini}
E.~Pavarini, S.~Biermann, A.~Poteryaev, A.I. Lichtenstein, A.~Georges, O.K.
  Andersen, Phys. Rev. Lett. \textbf{92}, 176403 (2004)

\bibitem{Vollhardt:Inoue94}
I.H. Inoue, I.~Hase, Y.~Aiura, A.~Fujimori, K.~Morikawa, T.~Mizokawa,
  Y.~Haruyama, T.~Maruyama, Y.~Nishihara, Physica C \textbf{235-240}, 1007
  (1994)

\bibitem{Vollhardt:Inoue03}
I.H. Inoue, private communication (2003)

\bibitem{Vollhardt:Lee85}
P.A. Lee, T.V. Ramakrishnan, Rev. Mod. Phys. \textbf{57}, 287 (1985)

\bibitem{Vollhardt:mott49}
N.F. Mott, Proc. Phys. Soc. London, Sect. A \textbf{62}, 415 (1949)

\bibitem{Vollhardt:Anderson58}
P.W. Anderson, Phys. Rev. \textbf{109}, 1492 (1958)

\bibitem{Vollhardt:Janis_a}
V.~Jani\v{s}, D.~Vollhardt, Phys. Rev. B \textbf{46}, 15712 (1992)

\bibitem{Vollhardt:Ulmke_aa}
V.~Jani\v{s}, M.~Ulmke, D.~Vollhardt, Europhysics Letters \textbf{24}, 287
  (1993)

\bibitem{Vollhardt:Ulmke_ab}
M.~Ulmke, V.~Jani\v{s}, D.~Vollhardt, Phys. Rev. B \textbf{51}, 10411 (1995)

\bibitem{Vollhardt:Dobro_a}
V.~Dobrosavljevi\'{c}, G.~Kotliar, Phys. Rev. B \textbf{50}, 1430 (1994)

\bibitem{Vollhardt:Dobro_b}
M.C.O. Aguiar, V.~Dobrosavljevi\'{c}, E.~Abrahams, G.~Kotliar, Phys. Rev. B
  \textbf{73}, 115117 (2006)

\bibitem{Vollhardt:Potthoff_a}
M.~Potthoff, M.~Balzer, Phys. Rev. B \textbf{75}, 125112 (2007)

\bibitem{Vollhardt:Janis92a}
V.~Jani\v{s}, D.~Vollhardt, Phys. Rev. B \textbf{46}, 15712 (1992)

\bibitem{Vollhardt:Dobrosavljevic97}
V.~Dobrosavljevi{\'c}, G.~Kotliar, Phys. Rev. Lett. \textbf{78}, 3943 (1997)

\bibitem{Vollhardt:Dobrosavljevic03}
V.~Dobrosavljevi{\'c}, A.A. Pastor, B.K. Nikoli{\'c}, Europhys. Lett.
  \textbf{62}, 76 (2003)

\bibitem{Vollhardt:fehske}
F.X. Bronold, A.~Alvermann, H.~Fehske, Phil. Mag. \textbf{84}, 637 (2004)

\bibitem{Vollhardt:Byczuk05}
K.~Byczuk, W.~Hofstetter, D.~Vollhardt, Phys. Rev. Lett. \textbf{94}, 056404
  (2005)

\bibitem{Vollhardt:Byczuk09_b}
K.~Byczuk, W.~Hofstetter, D.~Vollhardt, Phys. Rev. Lett. \textbf{102}, 146403
  (2009)

\bibitem{Vollhardt:Byczuk03_aa}
K.~Byczuk, M.~Ulmke, D.~Vollhardt, Phys. Rev. Lett. \textbf{90}, 196403 (2003)

\bibitem{Vollhardt:Byczuk03_ab}
K.~Byczuk, M.~Ulmke, Eur. Phys. J. B \textbf{45}, 449 (2005)

\bibitem{Vollhardt:Byczuk04_a}
K.~Byczuk, W.~Hofstetter, D.~Vollhardt, Phys. Rev. B \textbf{69}, 045112 (2004)

\bibitem{Vollhardt:Lombardo_a}
P.~Lombardo, R.~Hayn, G.I. Japaridze, Phys. Rev. B \textbf{74}, 085116 (2006)

\bibitem{Vollhardt:Potthoff_07}
M.~Potthoff, M.~Balzer, Phys. Rev. B \textbf{75}, 125112 (2007)

\bibitem{Vollhardt:Unjong_a}
U.~Yu, K.~Byczuk, D.~Vollhardt, Phys. Rev. Lett. \textbf{100}, 246401 (2008)

\bibitem{Vollhardt:bec}
R.J. Anglin, W.~Ketterle, Nature \textbf{416}, 211 (2002)

\bibitem{Vollhardt:Greiner02}
M.~Greiner, O.~Mandel, T.~Esslinger, T.W. H{\"a}nsch, I.~Bloch, Nature
  \textbf{415}, 39 (2002)

\bibitem{Vollhardt:lewenstein06}
M.~Lewenstein, A.~Sanpera, V.~Ahufinger, B.~Damski, A.~Sen, U.~Sen, Advances in
  Physics \textbf{56}, 243 (2007)

\bibitem{Vollhardt:Bloch07}
I.~Bloch, J.~Dalibard, W.~Zwerger, Rev. Mod. Phys. \textbf{80}, 885 (2008)

\bibitem{Vollhardt:jaksch98}
D.~Jaksch, C.~Bruder, J.I. Cirac, C.W. Gardiner, P.~Zoller, Phys. Rev. Lett.
  \textbf{81}, 3108 (1998)

\bibitem{Vollhardt:Isacsson05}
A.~Isacsson, M.C. Cha, K.~Sengupta, S.M. Girvin, Phys. Rev. B \textbf{72},
  184507 (2005)

\bibitem{Vollhardt:Huber07}
S.D. Huber, E.~Altman, H.P. B{\"u}chler, G.~Blatter, Phys. Rev. B \textbf{75},
  085106 (2007)

\bibitem{Vollhardt:micheli06}
A.~Micheli, G.K. Brennen, P.~Zoller, Nature \textbf{2}, 341 (2006)

\bibitem{Vollhardt:Fisher_a}
M.P.A. Fisher, P.B. Weichman, G.~Grinstein, D.S. Fisher, Phys. Rev. B
  \textbf{40}, 546 (1989)

\bibitem{Vollhardt:Byczuk08_a}
K.~Byczuk, D.~Vollhardt, Phys. Rev. B \textbf{77}, 235106 (2008)

\bibitem{Vollhardt:Snoek_a}
A.~Hubener, M.~Snoek, W.~Hofstetter, Phys. Rev. B \textbf{80}, 245109 (2009)

\bibitem{Vollhardt:Tong_a}
W.J. Hu, N.H. Tong, Phys. Rev. B \textbf{80}, 245110 (2009)

\bibitem{Vollhardt:Anders_a}
P.~Anders, E.~Gull, L.~Pollet, M.~Troyer, P.~Werner, Phys. Rev. Lett.
  \textbf{105}, 096402 (2010)

\bibitem{Vollhardt:Byczuk09_a}
K.~Byczuk, D.~Vollhardt, Ann. Phys. (Berlin) \textbf{18}, 622 (2009)

\bibitem{Vollhardt:Dziarmaga2010a}
J.~Dziarmaga, Adv. Phys. \textbf{59}, 1063 (2010)

\bibitem{Vollhardt:Polkovnikov2010a}
A.~Polkovnikov, K.~Sengupta, A.~Silva, M.~Vengalattore, Rev. Mod. Phys. \textbf{83}, 863 (2011)

\bibitem{Vollhardt:Perfetti2006}
L.~Perfetti, P.A. Loukakos, M.~Lisowski, U.~Bovensiepen, H.~Berger,
  S.~Biermann, P.S. Cornaglia, A.~Georges, M.~Wolf, Phys. Rev. Lett.
  \textbf{97}, 067402 (2006)

\bibitem{Vollhardt:Wall2009}
S.~Wall, D.~Brida, S.R. Clark, H.P. Ehrke, D.~Jaksch, A.~Ardavan, S.~Bonora,
  H.~Uemura, Y.~Takahashi, T.~Hasegawa, H.~Okamoto, G.~Cerullo, A.~Cavalleri,
  Nature Physics \textbf{7}, 114 (2011)

\bibitem{Vollhardt:Schmidt02}
P.~Schmidt, H.~Monien, arXiv:cond-mat/0202046

\bibitem{Vollhardt:Turkowski05}
V.~Turkowski, J.K. Freericks, Phys. Rev. B \textbf{71}, 085104 (2005)

\bibitem{Vollhardt:keldyshintro}
R.~van Leeuwen, N.E. Dahlen, G.~Stefanucci, C.O. Almbladh, U.~von Barth, in
  \emph{Time-dependent density functional theory}, \emph{Lecture Notes in
  Physics}, vol. 706, ed. by M.A.L. Marques, C.A. Ullrich, F.~Nogueira,
  A.~Rubio, K.~Burke, E.K.U. Gross (Springer, Berlin, 2006)

\bibitem{Vollhardt:Eckstein2010a}
M.~Eckstein, M.~Kollar, P.~Werner, Phys. Rev. B \textbf{81}, 115131 (2010)

\bibitem{Vollhardt:Eckstein2009}
M.~Eckstein, M.~Kollar, P.~Werner, Phys. Rev. Lett. \textbf{103}, 056403 (2009)

\bibitem{Vollhardt:Eckstein2010e}
M.~Eckstein, P.~Werner, Phys. Rev. B \textbf{82}, 115115 (2010)

\bibitem{Vollhardt:Eckstein2010epjst}
M.~Eckstein, A.~Hackl, S.~Kehrein, M.~Kollar, M.~Moeckel, P.~Werner, F.A. Wolf,
  Eur. Phys. J. Special Topics \textbf{180}, 217 (2010)

\bibitem{Vollhardt:Freericks09}
J.K. Freericks, H.R. Krishnamurthy, T.~Pruschke, Phys. Rev. Lett. \textbf{102},
  136401 (2009)

\bibitem{Vollhardt:Eckstein2008c}
M.~Eckstein, M.~Kollar, Phys. Rev. B \textbf{78}, 245113 (2008)

\bibitem{Vollhardt:Moritz2010a}
B.~Moritz, T.P. Devereaux, J.K. Freericks, Phys. Rev. B \textbf{81}, 165112
  (2010)

\bibitem{Vollhardt:Eckstein2008b}
M.~Eckstein, M.~Kollar, Phys. Rev. B \textbf{78}, 205119 (2008)

\bibitem{Vollhardt:Freericks06b}
J.K. Freericks, V.M. Turkowski, V.~Zlati{\'c}, Phys. Rev. Lett. \textbf{97},
  266408 (2006)

\bibitem{Vollhardt:Tran08x}
M.T. Tran, Phys. Rev. B \textbf{78}, 125103 (2008)

\bibitem{Vollhardt:Freericks08}
J.K. Freericks, Phys. Rev. B \textbf{77}, 075109 (2008)

\bibitem{Vollhardt:Joura2008a}
A.V. Joura, J.K. Freericks, T.~Pruschke, Phys. Rev. Lett. \textbf{101}, 196401
  (2008)

\bibitem{Vollhardt:Tsuji08}
N.~Tsuji, T.~Oka, H.~Aoki, Phys. Rev. B \textbf{78}, 235124 (2008)

\bibitem{Vollhardt:Tsuji2009a}
N.~Tsuji, T.~Oka, H.~Aoki, Phys. Rev. Lett. \textbf{103}, 047403 (2009)

\bibitem{Vollhardt:Eckstein2010d}
M.~Eckstein, T.~Oka, P.~Werner, Phys. Rev. Lett. \textbf{105}, 146404 (2010)

\bibitem{Vollhardt:Tsuji2010a}
N.~Tsuji, T.~Oka, P.~Werner, H.~Aoki, Phys. Rev. Lett. \textbf{106}, 236401 (2011)

\bibitem{Vollhardt:Eckstein2008a}
M.~Eckstein, M.~Kollar, Phys. Rev. Lett. \textbf{100}, 120404 (2008)

\bibitem{Vollhardt:EcksteinNJP2010}
M.~Eckstein, M.~Kollar, New J. Phys. \textbf{12}, 055012 (2010)

\bibitem{Vollhardt:Eurich2010a}
N.~Eurich, M.~Eckstein, P.~Werner, Phys. Rev. B \textbf{83}, 155122 (2011)

\bibitem{Vollhardt:Rigol2008a}
M.~Rigol, V.~Dunjko, M.~Olshanii, Nature \textbf{452}, 854 (2008)

\bibitem{Vollhardt:moeckel08}
M.~Moeckel, S.~Kehrein, Phys. Rev. Lett. \textbf{100}, 175702 (2008)

\bibitem{Vollhardt:moeckel09}
M.~Moeckel, S.~Kehrein, Ann. Phys. \textbf{\bf 324}, 2146 (2009)

\bibitem{Vollhardt:Greiner}
M.~Greiner, O.~Mandel, T.W. H\"ansch, I.~Bloch, Nature \textbf{419}, 51 (2002)

\bibitem{Vollhardt:Rapp}
A.~Rapp, G.~Zarand, C.~Honerkamp, W.~Hofstetter, Phys. Rev. Lett. \textbf{98},
  160405 (2007)

\bibitem{Vollhardt:Snoek}
M.~Snoek, I.~Titvinidze, C.~Toke, K.~Byczuk, W.~Hofstetter, New J. Phys.
  \textbf{10}, 093008 (2008)

\bibitem{Vollhardt:Mott2}
U.~Schneider, L.~Hackerm{\"u}ller, S.~Will, T.~Best, I.~Bloch, T.A. Costi, R.W.
  Helmes, D.~Rasch, A.~Rosch, Science \textbf{322}, 1520 (2008)

\bibitem{Vollhardt:Freericks-book}
J.K. Freericks, \emph{Transport in multilayered nanostructures --- {T}he
  dynamical mean-field approach} (Imperial College Press, London, 2006)

\bibitem{Vollhardt:Potthoff99}
M.~Potthoff, W.~Nolting, Phys. Rev. B \textbf{59}, 2549 (1999)

\bibitem{Vollhardt:Takizawa06}
M.~Takizawa, H.~Wadati, K.~Tanaka, M.~Hashimoto, T.~Yoshida, A.~Fujimori,
  A.~Chikamtsu, H.~Kumigashira, M.~Oshima, K.~Shibuya, T.~Mihara, T.~Ohnishi,
  M.~Lippmaa, M.~Kawasaki, H.~Koinuma, S.~Okamoto, A.J. Millis, Phys. Rev.
  Lett. \textbf{97}, 057601 (2006)

\bibitem{Vollhardt:Chen07}
L.~Chen, J.K. Freericks, Phys. Rev. B \textbf{75}, 1251141 (2007)

\bibitem{Vollhardt:Byczuk08x}
K.~Byczuk, in \emph{Condensed Matter Physics in the Prime of the 21st Century:
  {P}henomena, Materials, Ideas, Methods}, ed. by J.~Jedrzejewski (World
  Scientific, Singapore, 2008), p.~1

\bibitem{Vollhardt:Helmes08}
R.W. Helmes, T.A. Costi, A.~Rosch, Phys. Rev. Lett. \textbf{100}, 056403 (2008)

\end{thebibliography}
\end{document}